\pdfoutput=1
\documentclass[prd,reprint,twocolumn,superscriptaddress,letterpaper,nofootinbib,
               nopreprintnumbers,showpacs]{revtex4-1}

\usepackage{amsmath}
\usepackage{microtype}
\usepackage{siunitx}
\usepackage{txfonts}
\usepackage{verbatim}
\usepackage{dcolumn}

\usepackage[usenames, dvipsnames, svgnames, table]{xcolor}
\usepackage[colorlinks=true, citecolor=RoyalBlue,
            linkcolor=BrickRed, urlcolor=ForestGreen]{hyperref}
\usepackage{graphicx}
\usepackage[justification=justified]{caption}
\usepackage[justification=justified]{subcaption}
\usepackage{float}
\usepackage{lineno}
\usepackage{multirow}
\usepackage{placeins}

\usepackage{acronym}
\usepackage{ifthen}
\usepackage{svn-multi}

\newcommand*\patchAmsMathEnvironmentForLineno[1]{%
  \expandafter\let\csname old#1\expandafter\endcsname\csname #1\endcsname
  \expandafter\let\csname oldend#1\expandafter\endcsname\csname end#1\endcsname
  \renewenvironment{#1}%
     {\linenomath\csname old#1\endcsname}%
     {\csname oldend#1\endcsname\endlinenomath}}%
\newcommand*\patchBothAmsMathEnvironmentsForLineno[1]{%
  \patchAmsMathEnvironmentForLineno{#1}%
  \patchAmsMathEnvironmentForLineno{#1*}}%
\AtBeginDocument{%
\patchBothAmsMathEnvironmentsForLineno{equation}%
\patchBothAmsMathEnvironmentsForLineno{align}%
\patchBothAmsMathEnvironmentsForLineno{flalign}%
\patchBothAmsMathEnvironmentsForLineno{alignat}%
\patchBothAmsMathEnvironmentsForLineno{gather}%
\patchBothAmsMathEnvironmentsForLineno{multline}%
}

\begin{document}

\title{Reconstructing the calibrated strain signal in the Advanced LIGO detectors }
\author{A. D. Viets,${}^{1,2}$
M. Wade,${}^{3}$
A. L. Urban,${}^{4}$
S. Kandhasamy,${}^{5}$
J. Betzwieser,${}^{5}$
Duncan A. Brown,${}^{6}$
J. Burguet-Castell,
C. Cahillane,${}^{4}$
E. Goetz,${}^{7,8}$
K. Izumi,${}^{7}$
S. Karki,${}^{7,9}$
J. S. Kissel,${}^{7}$
G. Mendell,${}^{7}$
R. L. Savage,${}^{7}$
X. Siemens,${}^{1}$
D. Tuyenbayev,${}^{7,10}$
A. J. Weinstein${}^{4}$
}\noaffiliation
\affiliation{University of Wisconsin-Milwaukee, Milwaukee, WI 53201, USA}
\affiliation{Concordia University Wisconsin, Mequon, WI 53097, USA}
\affiliation{Kenyon College, Gambier, OH 43022, USA}
\affiliation{California Institute of Technology, Pasadena, CA 91125, USA}
\affiliation{LIGO Livingston Observatory, Livingston, LA 70803, USA}
\affiliation{Syracuse University, Syracuse, NY 13244, USA}
\affiliation{LIGO Hanford Observatory, Richland, WA 99352, USA}
\affiliation{University of Michigan, Ann Arbor, MI 48109, USA}
\affiliation{University of Oregon, Eugene, OR 97403, USA}
\affiliation{University of Texas Rio Grande Valley, Brownsville, TX 78520, USA}

\begin{abstract}

Advanced LIGO's raw detector output needs to be calibrated to compute dimensionless strain $h(t)$.   
Calibrated strain data is produced
in the time domain using both a low-latency, online procedure and a high-latency, offline procedure.
The low-latency $h(t)$ data stream is produced in two stages, the first of which is performed on the same
computers that operate the detector's feedback control system.  This stage, referred to as the front-end
calibration, uses infinite impulse response (IIR) filtering and performs
all operations at a 16384 Hz digital sampling rate.  Due to several limitations, this procedure currently introduces certain systematic
errors in the calibrated strain data, motivating the second stage of the low-latency procedure, known as the low-latency \texttt{gstlal} calibration pipeline.
The \texttt{gstlal} calibration pipeline uses finite impulse response (FIR) filtering to apply corrections to the output of the front-end calibration.  It applies time-dependent correction factors to the sensing and actuation components of the calibrated strain to reduce systematic errors.  The \texttt{gstlal} calibration pipeline is also used in high latency to recalibrate the data, which is necessary due mainly to online dropouts in the calibrated data and identified improvements to the calibration models or filters.

\end{abstract}


\maketitle

\section{Introduction}

Gravitational waves (GWs) represent a new messenger for astronomy, carrying information about compact objects
in the local universe such as neutron stars and black holes. To date, the Laser Interferometer Gravitational-wave
Observatory (LIGO) and the Virgo detector have observed several transient gravitational-wave signals from merging stellar-mass black hole binaries \cite{GW150914,
GW151226, O1BBHPaper, GW170104, GW170814} and a binary neutron star system \cite{GW170817},
and have recently finished the second observing run (O2) of the Advanced LIGO era. There are two LIGO observatories in
North America: one in Hanford, WA (H1) and another in Livingston, LA (L1).  The Virgo detector is located in Cascina, Italy.

\begin{figure}[!t]
    \centering
    \includegraphics[width=\columnwidth]{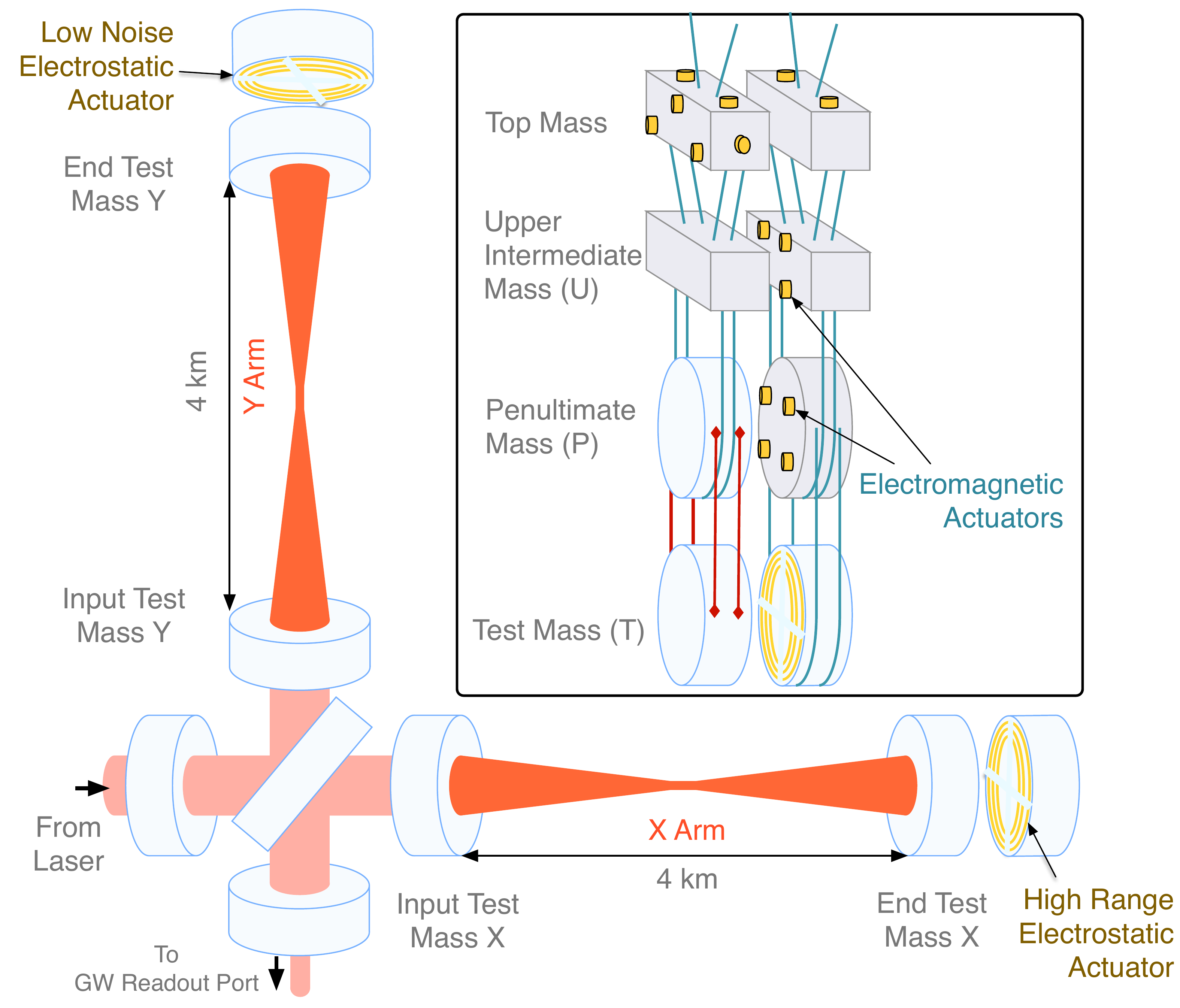}
    \caption{Simplified diagram of an Advanced LIGO detector. A set of four highly reflective mirrors (Input Test Mass Y, End Test Mass Y, Input Test Mass X, End Test Mass X)
act as test masses, forming an orthogonal pair of Fabry-P\'erot cavities. Laser light enters from the
lower left, passes through a beamsplitter, and enters resonant cavities in each 4 km arm. The laser light
then recombines out of phase at the gravitational-wave readout port, which records an error signal used to
actuate the end test masses to hold each cavity in resonance. For clarity, only the optics of the lowest suspension
stage are shown. \textit{Inset:} an illustration of one of the dual-chain, quadruple pendulum
suspension systems with actuators. This figure is reproduced with permission from \cite{GW150914CalPaper}.
\label{fig:aLIGO}}
\end{figure}

Each LIGO detector consists of two orthogonal arms, $L_{\rm x}$ and $L_{\rm y}$, roughly 4 km in length (Fig.
\ref{fig:aLIGO}). External gravitational-wave signals are measured from changes in the differential arm
(DARM) length,
\begin{equation}
\Delta L_{\rm free}(t) = \Delta L_{\rm x}(t) - \Delta L_{\rm y}(t),
\end{equation}
between pairs of test masses at opposite ends of $L_{\rm x}$ and $L_{\rm y}$.  Differential length
variations are measured using interferometric techniques \cite{aLIGOLSCPaper}: the test masses consist of highly reflective mirrors that
form a pair of resonant Fabry-P\'erot cavities. Input laser light passes through a beamsplitter, enters the
resonant cavities, then recombines out of phase at an output photodiode.  Any power fluctuations measured by the output photodiode
will then correspond to differential arm length changes.  To further improve sensitivity, a power recycling mirror at the input reflects laser light back into the arms of the interferometer to increase the laser power stored in the arms.  A signal recycling mirror at the output enhances the interferometer's sensitivity in the frequency band of interest \cite{GW150914DetectorPaper, AdvLIGOPaper, BuonnanoChen2002}.

However, we cannot measure $\Delta L_{\rm free}(t)$ directly because the differential arm length degree of freedom is held in
resonance by servos actuating the test mass positions in response to external stimuli, actively suppressing low-frequency fluctuations
in $\Delta L_{\rm free}(t)$ \cite{aLIGOLSCPaper}. The servos are part of a feedback control system, known as the DARM control loop, whose residual signal, known as the error signal, is our primary
observable, and from which any external length disturbance needs to be reconstructed. In practice, this is done by
modeling various transfer functions in the DARM control loop, usually with reference measurements taken prior
to the start of science-quality observing runs, and then applying them as filters in the time domain \cite{GW150914CalPaper, iLIGOhoft}.  

Calibration was originally done in the frequency domain.  The idea of producing a more convenient time-domain calibration came from GEO600 and has been employed since Initial LIGO's second science run \cite{iLIGOhoft}.  The time-domain calibration offers two important benefits over a frequency-domain calibration.  First, time-domain calibration can be provided with negligible latency, which is what we are striving to achieve, while frequency-domain calibration contains inherent latency.  Second, time-domain calibration is a more flexible data product that allows for any length Fourier transform that is needed or desired by data analysis efforts.

Absolute displacement calibration is achieved using auxiliary laser systems, called photon calibrators \cite{aLIGOPCALPaper}.   They induce fiducial test mass displacements via radiation pressure with force coefficients derived from laser power measurements traceable to SI units. The powers of these laser beams that reflect from the the test mass surfaces are measured using calibrated sensors, thus providing continuous absolute calibration when the interferometers are operating in their nominal configurations.  The overall 1-$\sigma$ uncertainty in the displacements induced by photon calibrators is 0.75\% \cite{aLIGOPCALPaper} with the long-term stability of the calibrated length variations verified during year-long observing runs \cite{P1100013}.  They thus also provide stable references for continuously monitoring temporal variations in the responses of the interferometers as described in Sec.~\ref{sec:introkappas}

Changes in the responses of the detectors occur due to, for example, slight drifts in the alignment of the optics.  These changes are small enough that they do not impact operation of the interferometers. However, if uncompensated, they would cause systematic biases in the calibrated strain data.  This would negatively impact estimation of parameters of astrophysical gravitational-wave sources \cite{Craig}. Fortunately, variations
known to have a significant impact, such as laser-power fluctuations in the Fabry-P\'erot cavities, are small and slow compared to the rapid timescale (16384 Hz) on which we record and analyze data.
To a good approximation, we can treat them as either time-dependent magnitudes of DARM loop transfer
functions, or changes in the zero or pole frequencies of those transfer functions \cite{GW150914CalPaper, CALTimeDependence}.

\begin{figure*}[!t]
    \centering
    \includegraphics[scale=1]{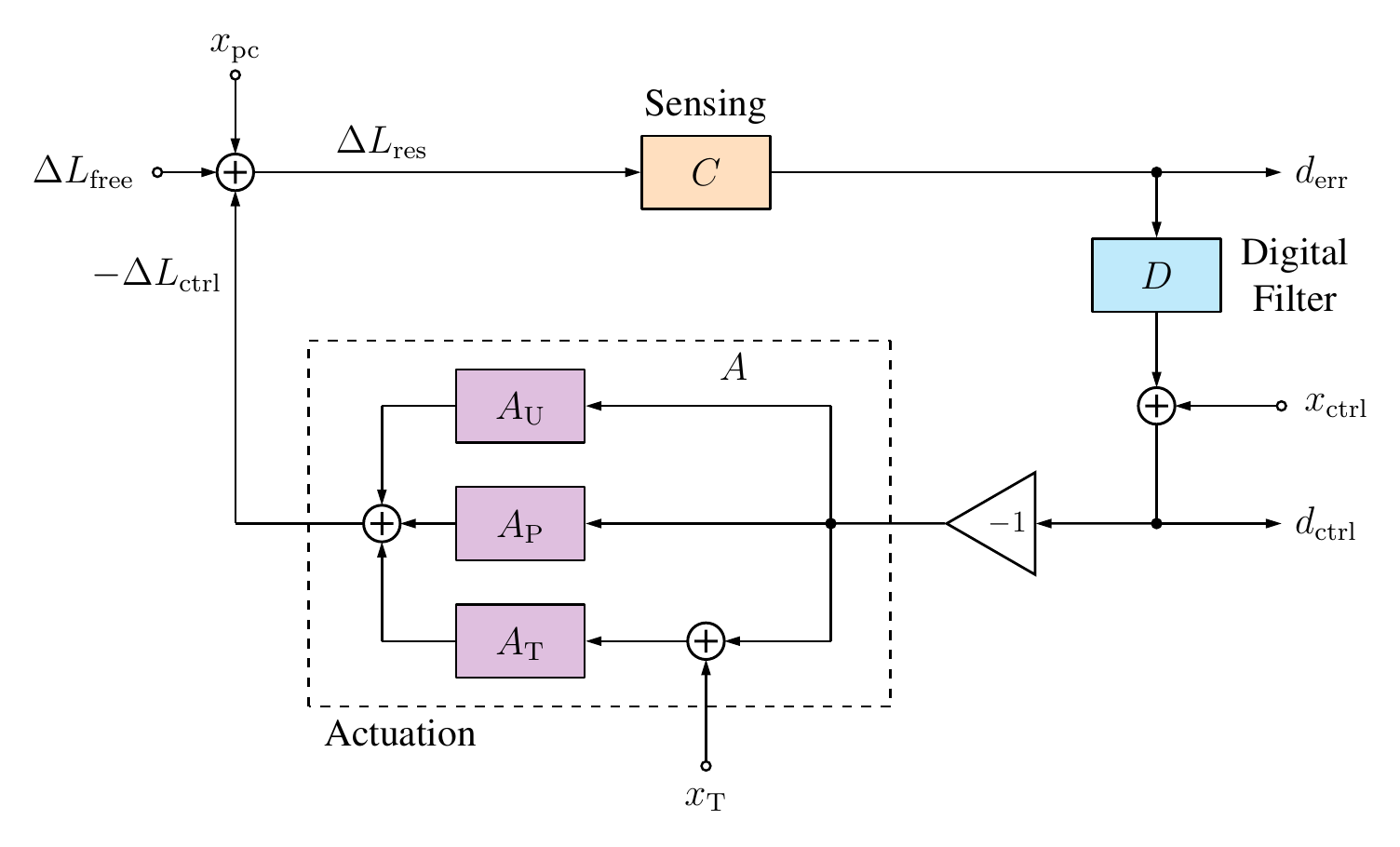}
    \caption{Block diagram of the Advanced LIGO differential arm (DARM) length feedback control loop.
Here, the filter $C$ is a sensing function representing the interferometer's response to changes in the measured differential
arm length; $D$ is a digital control filter; and the filters $A_{\rm U}$, $A_{\rm P}$
and $A_{\rm T}$ represent actuation force-to-length responses for the upper-intermediate
 (U), penultimate (P) and test mass (T) stages of LIGO's quadruple pendulum
suspension system. Length disturbances from external sources, including gravitational waves,
enter the loop as $\Delta L_{\rm free}$. The controlled length, $\Delta L_{\rm ctrl}$, is
subtracted from this to give a length change residual, $\Delta L_{\rm res}$, which passes
through the sensing function resulting in a digital error signal, $d_{\rm err}$. The error signal
is then filtered through a set of digital filters $D$ to create a digital control signal, $d_{\rm ctrl}$.  The digital filter $D$ is a linear combination of a low-pass filter and notch filters that prevent excitation of resonances in the test mass suspensions, among other things. The digital control signal $d_{\rm ctrl}$ is
used to actuate against the suspension pendula to hold the optical cavities in resonance.
Calibration of Advanced LIGO strain data consists of reconstructing $\Delta L_{\rm free}$
from $d_{\rm err}$, $d_{\rm ctrl}$ and models of the filters $C^{-1}$ and $A = A_{\rm U}
+ A_{\rm P} + A_{\rm T}$. Lastly, to check accuracy and precision we inject known sinusoidal
excitations (i.e. calibration lines) at three points in the loop, represented by
$x_{\rm pc}$ (injected through radiation pressure by a photon calibrator), $x_{\rm ctrl}$
(injected into the control signal), and $x_{\rm T}$ (injected into the test mass actuation
stage). \label{fig:DARM_loop}}
\end{figure*}

A simplified model of the DARM control loop is shown in Fig. \ref{fig:DARM_loop}. Any length disturbances
that arise from external sources, including gravitational waves and displacement noise, enter the loop as
$\Delta L_{\rm free}(t)$. A controlled differential length, $\Delta L_{\rm ctrl}(t)$, compensates part
of this external displacement, giving a residual displacement
\begin{equation}\label{eq:DeltaL}
\Delta L_{\rm res}(t) = \Delta L_{\rm free}(t) - \Delta L_{\rm ctrl}(t).
\end{equation}
The residual is directly sensed by the detector and read out as an error signal
\begin{equation}\label{eq:d_err}
d_{\rm err}(t) = C \ast \Delta L_{\rm res}(t)
\end{equation}
where $C$ is a sensing function representing the opto-mechanical response of the interferometer to changes in the DARM length.  The operation
\begin{equation}\label{eq:convolution}
F \ast g(t) = \int_{-\infty}^{\infty} F(\tau) \, g(t - \tau) \,\, d\tau
\end{equation}
denotes convolution of some time domain filter $F$ with some signal $g(t)$.  As indicated by Eq.~\eqref{eq:d_err}, our model relies on the linearity of the detector's response, which is enforced by holding the differential degree of freedom slightly off anti-resonance.  This is done by digitally adding a very small DC offset, and causes a small amount of light to exit the interferometer at the readout port. \cite{EvanHall}

The error signal is also used to create a control signal, $d_{\rm ctrl}(t)$.  The control signal feeds into the actuation system, which acts on the suspension pendula to hold the Fabry-P\'erot cavities in resonance (Fig. \ref{fig:DARM_loop}). The controlled differential length is
\begin{equation}\label{eq:d_ctrl}
\Delta L_{\rm ctrl} (t) = A \ast d_{\rm ctrl}(t)
\end{equation}
where $A$ is a filter representing the combined electromechanical actuation response from each pendulum stage. In
terms of the observable $d_{\rm err}(t)$ and its corollary $d_{\rm ctrl}(t)$, the external differential length is therefore
\begin{equation}\label{eq:hoft}
\Delta L_{\rm free}(t) = C^{-1} \ast d_{\rm err}(t) + A \ast d_{\rm ctrl}(t)
\end{equation}
where $C^{-1}$ is the filter inverse of $C$.

The external differential length can also be computed from only the error signal using the response function that relates the DARM length to the error signal: $\widetilde{\Delta L}_{\rm free}(f) = \tilde{R}(f) \tilde{d}_{\rm err}(f)$, where the tildes denote Fourier transformation into the frequency domain.  The response function is given by
\begin{equation}\label{eq:responseFunction}
    \tilde{R}(f) = \frac{1 + \tilde{A}(f) \tilde{D}(f) \tilde{C}(f)}{\tilde{C}(f)} \, .
\end{equation}
Historically, we have used both the error and control signals when reconstructing $\Delta L_{\rm free}$ \cite{iLIGOhoft}, as shown in Eq.~\eqref{eq:hoft}, for two main reasons.  First, this allows the computation of $\Delta L_{\rm free}(t)$ to be insensitive to the digital filters $D$, which are often changed to improve the control loop design.\footnote{Now that time dependent correction factors are included in the calibration process, this is actually no longer true.  However, it remains a historical motivation for how methods were developed.}  Second, the predetermined control signal $x_{\rm ctrl}(t)$, shown in Fig.~\ref{fig:DARM_loop}, will not appear in the resulting $\Delta L_{\rm free}(t)$ spectrum due to its location in the feedback loop between the $d_{\rm err}(t)$ and $d_{\rm ctrl}(t)$ pickoff points.

The dimensionless strain $h(t)$ used for detecting gravitational
waves is a time series derived from $\Delta L_{\rm free}(t)$,
\begin{equation}
h(t) = \frac{\Delta L_{\rm free}(t)}{L},
\end{equation}
where $L = (L_x + L_y)/2$ is the average measured arm length. Since fluctuations in $L$ are negligible, the process of calibration therefore amounts to reconstructing $\Delta L_{\rm free}(t)$ from $d_{\rm err}(t)$ and $d_{\rm ctrl}(t)$.  Hereafter we will use $\Delta L_{\rm free}(t)$ or $h(t)$ interchangeably to refer to the final calibrated data product.


Eq. (\ref{eq:hoft}) establishes a method for reconstructing calibrated strain data from the error and control signals.  
We construct accurate physical models for the filters $C^{-1}$ and $A$ (then divide by
$L$, which does not change). In Sec.~\ref{sec:Overview}, we outline the important physical properties
of transfer functions associated with these filters and describe how we correct for small variations over time. A detailed
discussion of the full interferometric response, including its effect on calibration uncertainty, can be found elsewhere
\cite{GW150914CalPaper, CALTimeDependence, Craig}.  In Sec.~\ref{sec:pipelines}, we describe the software pipelines used to perform the calibration in Advanced LIGO's first and second observing runs.  In Sec.~\ref{sec:futureDevelopment} we discuss plans for future development, and in Sec.~\ref{sec:conclusion} we conclude.

\section{Frequency Domain Models}\label{sec:Overview}

\begin{figure*}[!t]
	\centering
	\includegraphics[width=\textwidth]{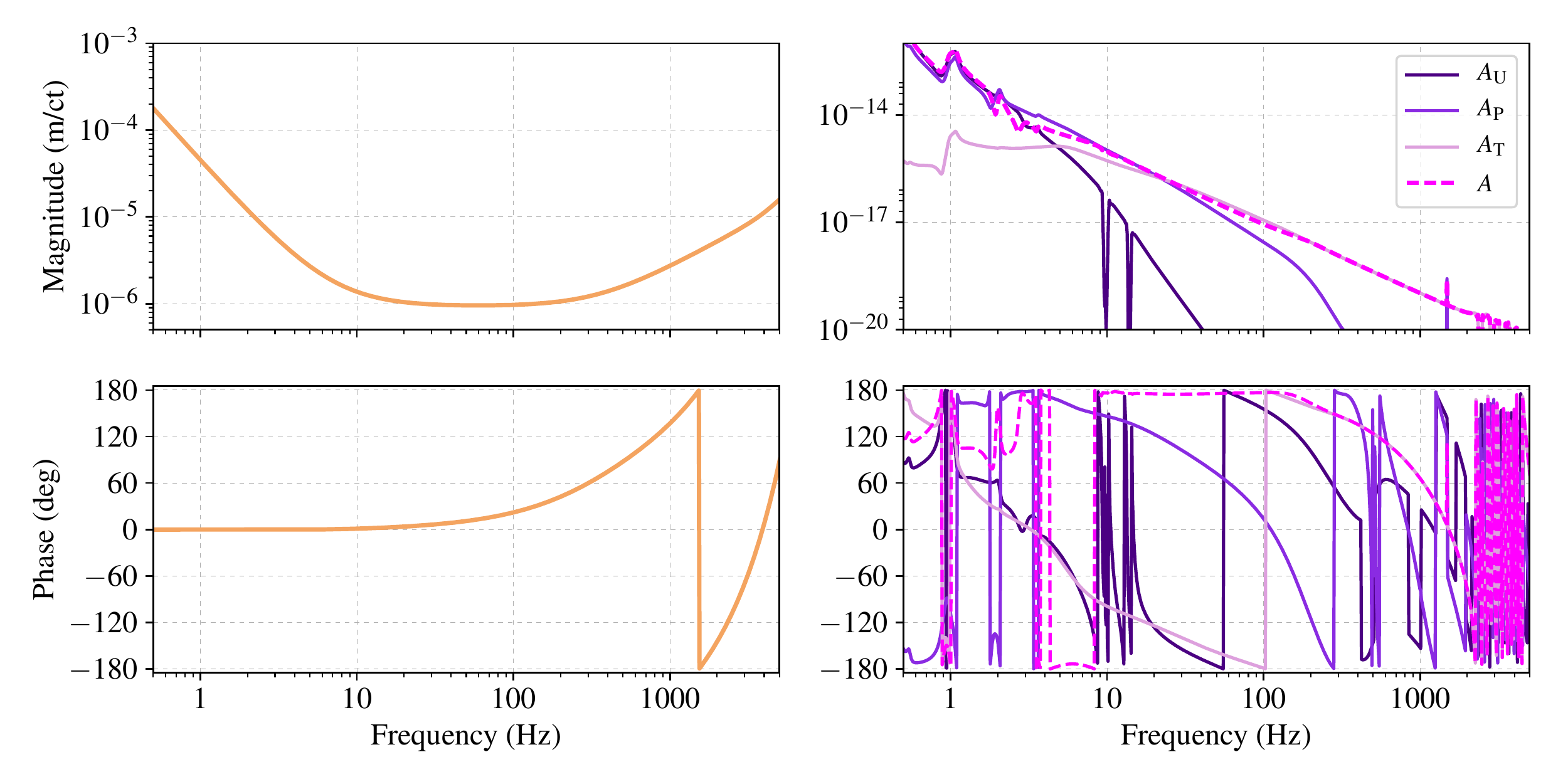}
	\caption{Frequency-domain models corresponding to the inverse sensing ($C^{-1}$, \textit{left}) and actuation
(\textit{right}) transfer functions as measured at LIGO Hanford Observatory (H1). The inverse sensing function
converts $d_{\rm err}$ counts to residual displacement, while the actuation function converts $d_{\rm ctrl}$ counts
to controlled displacement. Note, the factor of $\sim$10$^{9}$ difference between inverse sensing and total
actuation around 10 Hz is due to a very large gain applied to $d_{\rm err}$ to create $d_{\rm ctrl}$. At low frequency
the inverse sensing is dominated by a low-pass filter around 8 Hz,
while above $\sim$340 Hz it is affected by $f_{\rm cc}$ and above $\sim$1 kHz by analog-to-digital
conversion. In the actuation response, there are several notch filters used to avoid mechanical resonances. Above
$\sim$30 Hz, only the test mass stage (T) is actuated against. The corresponding curves at LIGO Livingston Observatory (L1) are
qualitatively similar, but differ slightly in scale and frequency dependence; see e.g. \cite{GW150914CalPaper}.
\label{fig:transfer_fncs}}
\end{figure*}

\subsection{Sensing Function}\label{ssec:Sensing}

After $\Delta L_{\rm ctrl}$ is subtracted from $\Delta L_{\rm free}$, the sensing function $C$ converts any residual test mass displacement $\Delta L_{\rm res}$ into
a digitized error signal representing the remaining fluctuations in laser power at the output photodiode.  This signal is measured as
$d_{\rm err}$ at the gravitational-wave readout port (Fig. \ref{fig:aLIGO}) with a sample rate of 16384 Hz. The
sensing function also accounts for responses of photodiodes and their analog electronics, light-travel time through
the Fabry-P\'erot cavities, and digital time delays and downsampling filters.

A standard suite of calibration measurements, as discussed in \cite{aLIGOPCALPaper, GW150914CalPaper}, informs a static reference model for the sensing function.  The static reference model is represented by 
\begin{eqnarray}\label{eq:Cstatic}
\tilde{C}^{\rm static}(f) &=&  \left(\frac{H_C}{1 + if/f_{\rm cc}}\right) \\
\nonumber
&& \times \left(\frac{f^2}{f^2 + f_{\rm s}^2 - i f f_{\rm s} / Q}\right)  \\
\nonumber
&& \times \ C_{\rm R}(f) \, \rm{exp}\left[{-2\pi \it{i} f \tau_{\rm C}}\right].
\end{eqnarray}
The gain $H_C$ gives the number of digital counts in $d_{\rm err}$ per unit differential length change, in meters. The pole $f_{\rm cc}$ is called the coupled cavity pole frequency;
it represents the characteristic frequency beyond which the detector response to gravitational waves is significantly
attenuated due to finite average photon storage time in the Fabry-P\'erot cavities. Both LIGO interferometers are designed
to have the same coupled cavity pole frequencies, but they differ due to differing losses in the optical
cavities.  During O2, the model reference values were $f_{\rm cc} = 360.0$ Hz at H1 and $f_{\rm cc} = 376.0$ Hz at L1.  $f_{\rm s}$ is the resonant frequency of the optical spring of the signal recycling cavity (SRC), with a model reference value of 6.91 Hz at H1 and 0 Hz at L1 during O2.  An optical spring exists in an opto-mechanical cavity if there is a linear relationship between the length of the cavity and the radiation pressure on the mirrors.  $Q$ is the quality factor of the SRC.  The next parameter in Eq. \eqref{eq:Cstatic}, $\tau_{\rm C}$, combines computational delay in acquiring the digital signal, the light-travel time across the length of each arm, and
a time advance of $\sim$\SI{11.7}{\micro\second} necessary to approximate the majority of the detector response with a single pole.  With this advance, the single-pole approximation is negligibly close to full analytical model across the relevant frequency band.  The last factor, $C_{\rm R}(f)$, encodes the remaining frequency dependence
above $\sim$1 kHz due to photodiode electronics and analog and digital signal processing filters. All parameters
are measured before the start of data collection by actuating the test masses with a swept-sine
signal, injected through a combination of the photon calibrator ($x_{\rm pc}$), the control signal ($x_{\rm ctrl}$), and the test mass actuation stage ($x_{\rm T}$), shown in Fig \ref{fig:DARM_loop}. These measurements are made roughly once every
few months and form a set of reference parameters for the full calibration model \cite{GW150914CalPaper}.

Parameters of the sensing function -- specifically, the gain $H_C$, the coupled cavity pole frequency $f_{\rm cc}$, the spring frequency $f_{\rm s}$, and the quality factor $Q$ of the SRC -- have been found to vary slowly with time.  These factors are all found to change on a timescale of $\sim$~minutes, which is slow
compared to the timescale of recording data ($\sim$~10$^{-4}$~s).  The full time-dependent sensing function model is 
\begin{eqnarray}\label{eq:C}
\tilde{C}(f; t) &=& \kappa_{\rm C}(t)  \left(\frac{H_{\rm C}}{1 + if/f_{\rm cc}(t)}\right) \\
\nonumber
&& \times \left(\frac{f^2}{f^2 + f_{\rm s}(t)^2 - i f f_{\rm s}(t) / Q(t)}\right)  \\
\nonumber
&& \times \ C_{\rm R}(f) \, \rm{exp}\left[{-2\pi \it{if}\tau_{\rm C}}\right] \ .
\end{eqnarray}
Variations in the gain $H_{\rm C}$ due to changes in laser power and alignment are captured by the factor $\kappa_{\rm C}(t)$, which is set to a value of 1
in the reference model and fluctuates by less than 10\% with time.

The reference model for $C^{-1}$, which converts power variations sensed at the anti-symmetric port of the interferometer to differential arm length variations and is
represented in the frequency domain by $1/\tilde{C}^{\rm static}(f)$, is plotted in Fig. \ref{fig:transfer_fncs}.

\subsection{Actuation Function}\label{ssec:Actuation}

The differential arm length is controlled using electromagnetic actuators on the top three pendulum stages and an electrostatic drive actuator on the bottom stage (Fig. \ref{fig:aLIGO}). While the topmost stage is connected to a seismic isolation system,
its actuation strength is relatively small in the Advanced LIGO sensitive frequency band, and the DARM feedback controls are not sent to this topmost stage.  The lowest three stages -- referred to as the upper
intermediate (U), penultimate (P), and test (T) mass stages -- are dominant above 10 Hz, and all these stages
are displaced in concert \cite{Suspensions2002, Suspensions2012, QUADSensorsAndActuators, aLIGOSEI}. The control signal ($d_{\rm ctrl}$)
is distributed to each stage in parallel, converted from digital counts to force, and used to actuate against the reaction mass at that stage. The
net result of this is the controlled length differential, $\Delta L_{\rm ctrl}$ (Fig. \ref{fig:DARM_loop}).

As with the sensing function, calibration measurements inform a static model for the actuation function.  The static reference model for the actuation is represented by a counts-to-length transfer function,
\begin{eqnarray}\label{eq:Astatic}
\tilde{A}^{\rm static}(f) &=& \left[\tilde{A}_{\rm U}(f) +
\tilde{A}_{\rm P}(f) + \tilde{A}_{\rm T}(f) \right]  \exp\left[-2\pi if\tau_{\rm A}\right] 
\end{eqnarray}
where $\tilde{A}_i(f)$ represents the frequency response of the $i$th
suspension stage actuator. Filtering that converts digital counts to actuation strength and splits the frequency content of $d_{\rm ctrl}$ across each stage of the actuation is folded into these functions
for brevity.  The low frequency content of $d_{\rm ctrl}$ is directed to the higher stages of the actuation system and the high frequency content is directed to the lower stages.  The computational time delay $\tau_{\rm A}$ accounts for digital-to-analog
conversion.  The modeled actuation transfer functions are plotted in Fig. \ref{fig:transfer_fncs}.  

The actuation function also has time-dependent gains for each stage.  The fully time-dependent actuation function model is
\begin{eqnarray}\label{eq:A}
\tilde{A}(f;t) &=& \left[ \kappa_{\rm U}(t) \tilde{A}_{\rm U}(f) +
\kappa_{\rm P}(t) \tilde{A}_{\rm P}(f) + \kappa_{\rm T}(t) \tilde{A}_{\rm T}(f) \right] \\
\nonumber
&& \times \exp\left[-2\pi if\tau_{\rm A}\right] \ .
\end{eqnarray}
The frequency-independent factors $\kappa_{\rm U}(t)$, $\kappa_{\rm P}(t)$, and $\kappa_{\rm T}(t)$ represent slowly-varying scalar fluctuations over time.  Changes are typically at the level of a few percent and vary on the timescale of $\sim$ hours.  $\kappa_{\rm T}$ represents the time-dependence of the electrostatic drive at the test mass stage of the actuation chain.  Physically, this time variation is understood to be due to charge build-up on the electrostatic drive.  The effect of $\kappa_{\rm P}$ and $\kappa_{\rm U}$ has so far only been tracked as a combined effect $\kappa_{\rm PU}$, which we have observed to vary very little with time.  Values of $\kappa_{\rm PU}$ that stray from unity are understood to originate from errors present in the model and occasional small changes in computational delays in the DARM feedback loop.

\begin{figure}
    \begin{center}
    \includegraphics[width=\columnwidth] {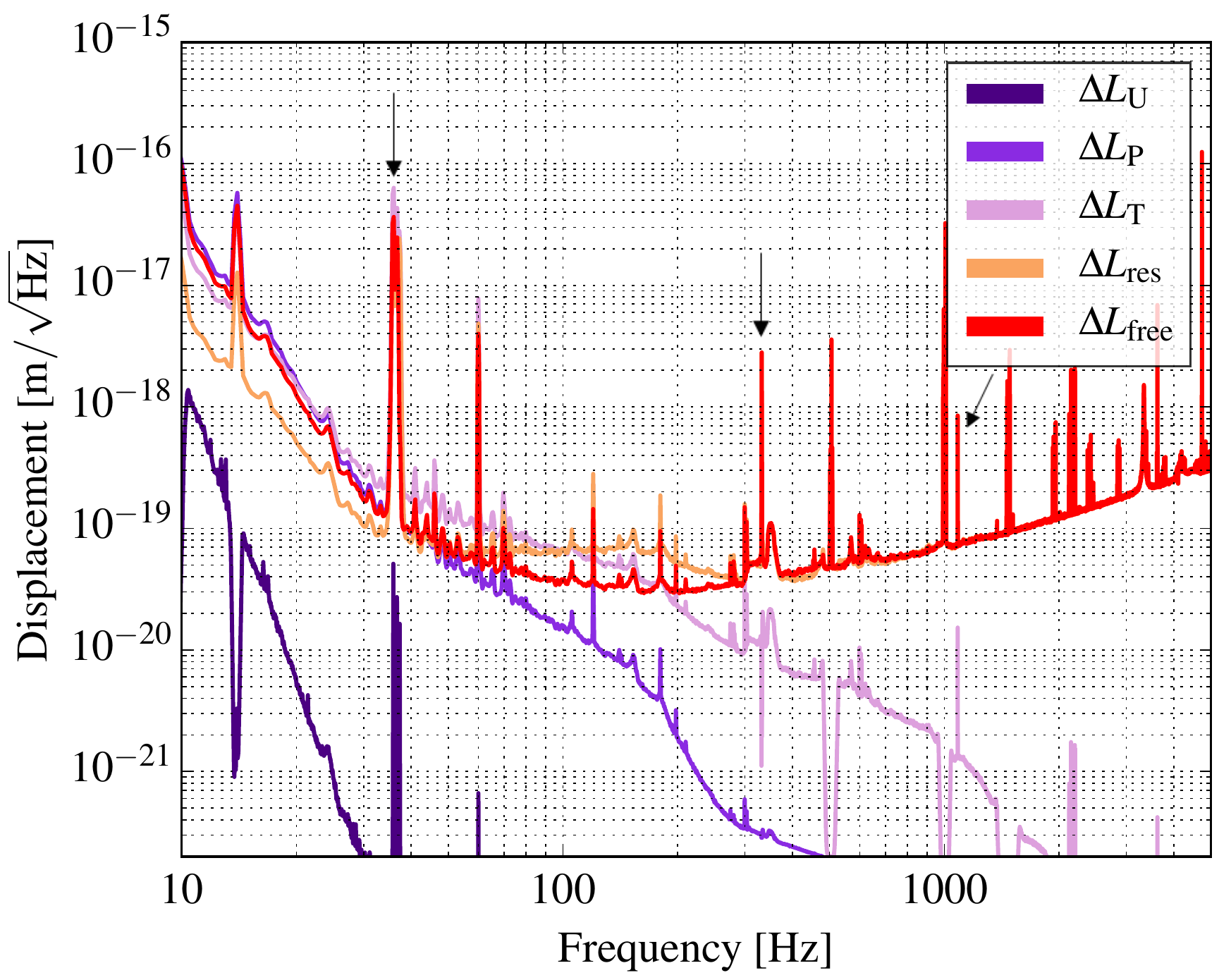}
    \end{center}
    \caption{Amplitude spectral densities for $\Delta L_{\rm res} = C^{-1} \ast d_{\rm err}(t)$, $\Delta L_{\rm T} = A_{\rm T} \ast d_{\rm ctrl}(t)$, $\Delta L_{\rm P} = A_{\rm P} \ast d_{\rm ctrl}(t)$, $\Delta L_{\rm U} = A_{\rm U} \ast d_{\rm ctrl}(t)$, and $\Delta L_{\rm free}$ across the relevant frequency band at H1. These were obtained as intermediate products from the high-latency \texttt{gstlal} calibration pipeline (Sec.~\ref{ssec:DCS}). Note the visible calibration lines, indicated with arrows: $f_{\rm T} = \SI{35.9}{\hertz}$ is injected using the electrostatic drive, and $f_1^{\rm pc} = \SI{36.7}{\hertz}$, $f_2^{\rm pc} = \SI{331.9}{\hertz}$, and $f_3^{\rm pc} = \SI{1083.7}{\hertz}$ are injected using the photon calibrator.}
    \label{fig:spectrumBreakdown}
\end{figure}

Fig.~\ref{fig:spectrumBreakdown} shows the relative strength of each component of the calibration across the aLIGO sensitive frequency range.  The contribution of the actuation function to the final calibration diminishes above several hundred Hertz, after which the calibration is dominated by the sensing function.  In the 30 - 300 Hz range, the actuation and sensing both contribute significantly to $\Delta L_{\rm free}$.


\subsection{Tracking and Compensating for Small Variations}\label{sec:introkappas}

\begin{figure*}[p]
	\centering
	\includegraphics[width=\textwidth]{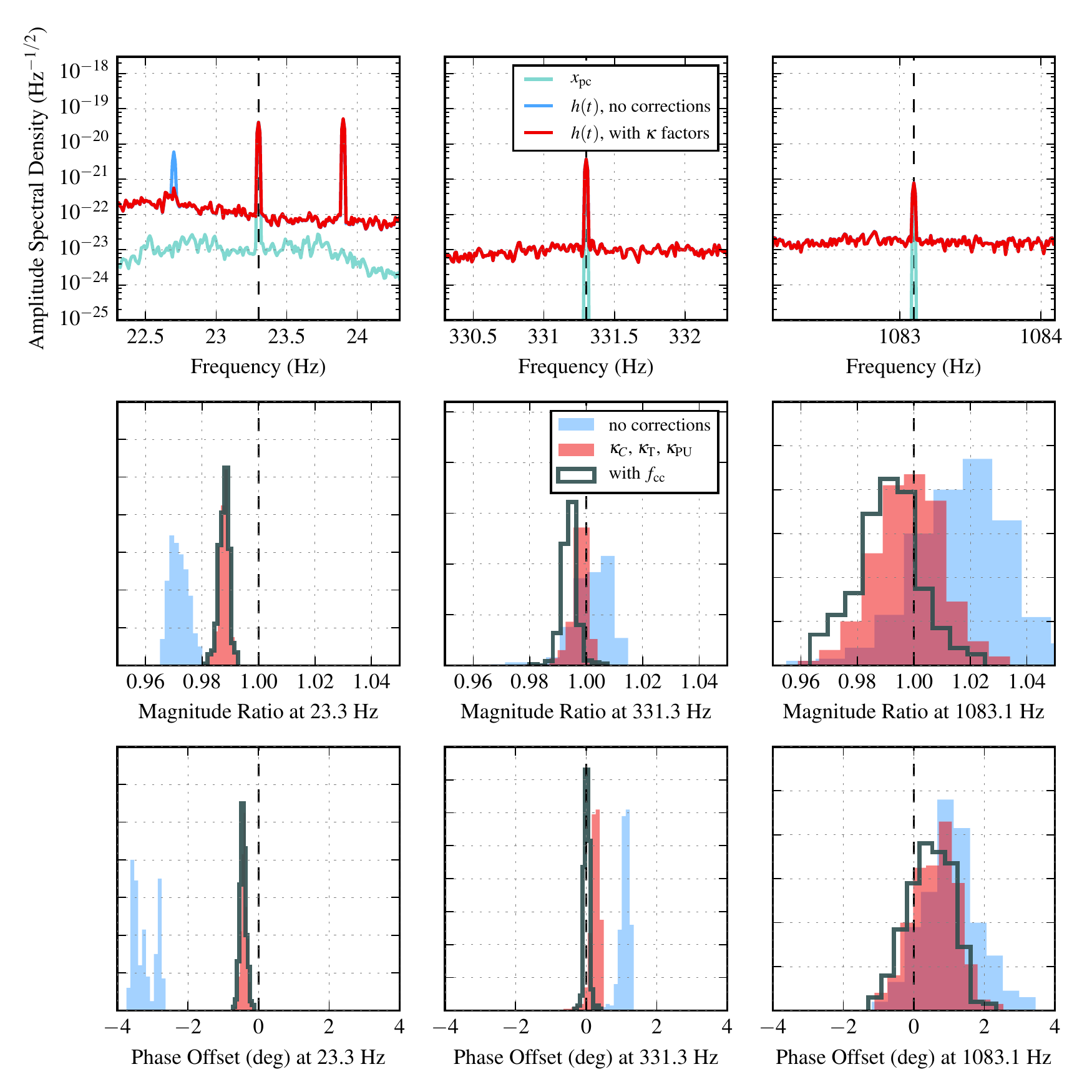}
	\caption{Impact on calibration accuracy from small deviations in the L1 detector over time. Data shown were collected between GPS seconds 1167536722 and 1167695629, on January 4 and 5, 2017. These deviations are quantified
by scalar time-dependent factors $\kappa_{\rm C}(t)$, $\kappa_{\rm T}(t)$, and $\kappa_{\rm PU}(t)$, which are applied to the time domain inverse sensing
and actuation filters. In addition, the time-dependence of the coupled cavity pole frequency $f_{\rm cc}$ is tracked but
not compensated for.  In the H1 detector, the optical spring frequency $f_{\rm s}$ and quality factor $Q$ of the SRC are also tracked but not compensated for.  \textit{Top:} Amplitude spectral densities of $h(t)$ near the calibration line frequencies. Three of these lines -- which live
at low, mid and high frequencies, respectively -- are injected through radiation pressure against the $y$-end test mass using a photon calibrator
($x_{\rm pc}$). \textit{Middle:} Histograms of the magnitude ratio $\vert\tilde{h}(f)\vert/\vert\tilde{x}_{\rm pc}(f)\vert$,
measured in the frequency domain at each $x_{\rm pc}$ line. A series of 100-second Fourier transforms of $h(t)$ and
$x_{\rm pc}(t)$ were computed covering 72 hours of L1 data surrounding the binary black hole merger signal GW170104 \cite{GW170104}.
Note that uncorrected strain contains 1-2\% systematic errors in magnitude at all three $x_{\rm pc}$ lines, which is
addressed by compensating for TDCFs.  The scale of random errors in these measurements is set by the
signal-to-noise ratio of the injected signal.  However, correcting for $f_{\rm cc}$ would introduce an overall $\sim$1\%
magnitude offset at mid and high frequencies due to systematics in the measurement of the sensing and actuation functions \cite{Craig}.
\textit{Bottom:} Histograms of the phase offset between $\tilde{x}_{\rm pc}$ and $\tilde{h}$ at each
$x_{\rm pc}$ line over the same period. Note that correcting for TDCFs and $f_{\rm cc}$ completely removes the
systematic error in phase at 331.3 Hz, which is in Advanced LIGO's most sensitive band. Results from H1 are similar, although
its calibration lines are at slightly different frequencies. During O2, there was also a larger variation in $f_{\rm cc}$ at H1
compared to L1. \label{fig:pcal_lines}} \end{figure*}

The calibration models vary in time as the detectors fluctuate and settle throughout operation.  While in Initial LIGO the only time-dependent correction applied to the calibration was the gain of the sensing function \cite{iLIGOhoft}, compensation for temporal variations is more complex in Advanced LIGO.  The time-dependent correction factors (TDCFs) for the current instruments include $\kappa_{\rm C}$, $f_{\rm cc}$, $\kappa_{\rm T}$, $\kappa_{\rm PU}$, $f_{\rm s}$, and $Q$ \cite{CALTimeDependence}.  In the H1 detector, the optical spring frequency $f_{\rm s}$ and quality factor $Q$ of the SRC are tracked, but at L1, $f_{\rm s} \approx \SI{3}{\hertz}$, which is below the sensitive frequency range of the Advanced LIGO detectors, and so the time-dependence of SRC detuning parameters is not currently tracked at L1. 


The TDCFs are measured by injecting loud, known sinusoidal excitations (i.e. calibration lines) and tracking their amplitude and phase in $d_{\rm err}$ over time.  While the calibration line placement varies in frequency at H1 compared to L1, calibration
lines at both detectors serve similar purposes.
Fig.~\ref{fig:DARM_loop} shows the injection point of each calibration line.  Length excitations due to all injected signals
are suppressed by the DARM control loop, and their influence on $d_{\rm err}$ can be predicted with models for $C$, $D$, and $A$.
Appendix~\ref{appendix:appendixA} discusses how each TDCF is computed using the calibration lines.  

Calibration accuracy can be quantified using the calibration lines injected with the photon calibrator, through comparison of their amplitude and phase as seen in the photon calibrator injection channel and in $h(t)$.
Amplitude spectral densities (ASDs) of $h(t)$ near each photon calibrator line frequency are shown in Fig.~\ref{fig:pcal_lines} to demonstrate the effect of correcting for small detector variations.
The data shown are taken from the L1 detector over a period of 72 hours surrounding binary black hole
merger GW170104 \cite{GW170104}. Consecutive Fourier transforms were used to compute strain data in the frequency
domain, using a relatively long integration time of 100 seconds in order to average out any short-duration transient signals (which might otherwise contaminate the measured line height)\footnote{Transient gravitational wave signals could be events like GW170104, which swept from 30 Hz to 500 Hz in
less than a second. These are signals that only last in our detector for a short period of
time but can in general span a wide range of frequencies. The sinusoidal injections, by
contrast, only exist at a single frequency but last throughout the observing run.} and to give a frequency resolution of 0.01
Hz. Without compensating for time variations, there can be up to a 15\% systematic error in magnitude and
1-4$^{\circ}$ systematic error in phase across Advanced LIGO's sensitive frequency range, which can impact parameter estimation \cite{RatesPaper, O1Stoch2017a, O1Stoch2017b, O1CW2017a, O1CW2017b, TestingGR2016, TestingGR2017, GW150914GRTests, Chernoff1993, O1BBHPaper, GW150914PEPaper}.  Improved calibration is obtained by compensating for these time-varying parameters.  As with taking reference measurements of the sensing and actuation functions, accurate tracking of the TDCFs during observation relies on the accuracy of the photon calibrator.

Having sketched the computations required for generation of a precisely calibrated $h(t)$ data stream, we now discuss how this is accomplished in Advanced LIGO.

\section{Calibration Pipelines} \label{sec:pipelines}

\begin{table*}[t]
\centering
\begin{tabular}{ ||c | c | c | c | c | c ||}
\hline
 \textbf{Pipeline} & \textbf{Latency} & \textbf{Accuracy }& \textbf{Input} & \textbf{Type of filtering} & \textbf{Purpose} \\
 \hline
 \hline
front-end (0) & real-time & $\lessapprox$ 14\% & $d_{\rm err}$ and $d_{\rm ctrl}$ & IIR & immediate commissioning feedback \\
\hline
low-latency \texttt{gstlal} (1) & $\mathcal{O}$(10 s) & $\lessapprox$ 2\% & output of front-end ($\Delta L_{\rm res}^0$ and $\Delta L_{\rm ctrl}^0$) & FIR & low-latency searches \\
\hline
high-latency \texttt{gstlal} (2) & $\mathcal{O}$(weeks) & n/a & $d_{\rm err}$ and $d_{\rm ctrl}$ & FIR & final science results \\
\hline
\end{tabular}
\caption{\label{table:pipelines} Summary of the significant differences and inherent purpose of the different calibration pipelines.  The number in parentheses next to each pipeline name indicates the numerical shorthand assigned to that pipeline.  For example $A^0$ indicates that this is the actuation function used by the front-end pipeline and $\Delta L_{\rm free}^0$ is the calibrated $\Delta L_{\rm free}$ reconstructed by the front-end pipeline.  The accuracy listed is the additional systematic error in magnitude when compared to the high-latency \texttt{gstlal} calibration pipeline, which produces the most accuracy $h(t)$ we are capable of producing at the time of writing.  This is not the total calibration uncertainty for the $h(t)$ produced by each pipeline.  A full study of the total calibration uncertainty can be found in \cite{Craig}.}
\end{table*}

\begin{figure*}[t]
\begin{center}
\includegraphics[width=0.5\textwidth] {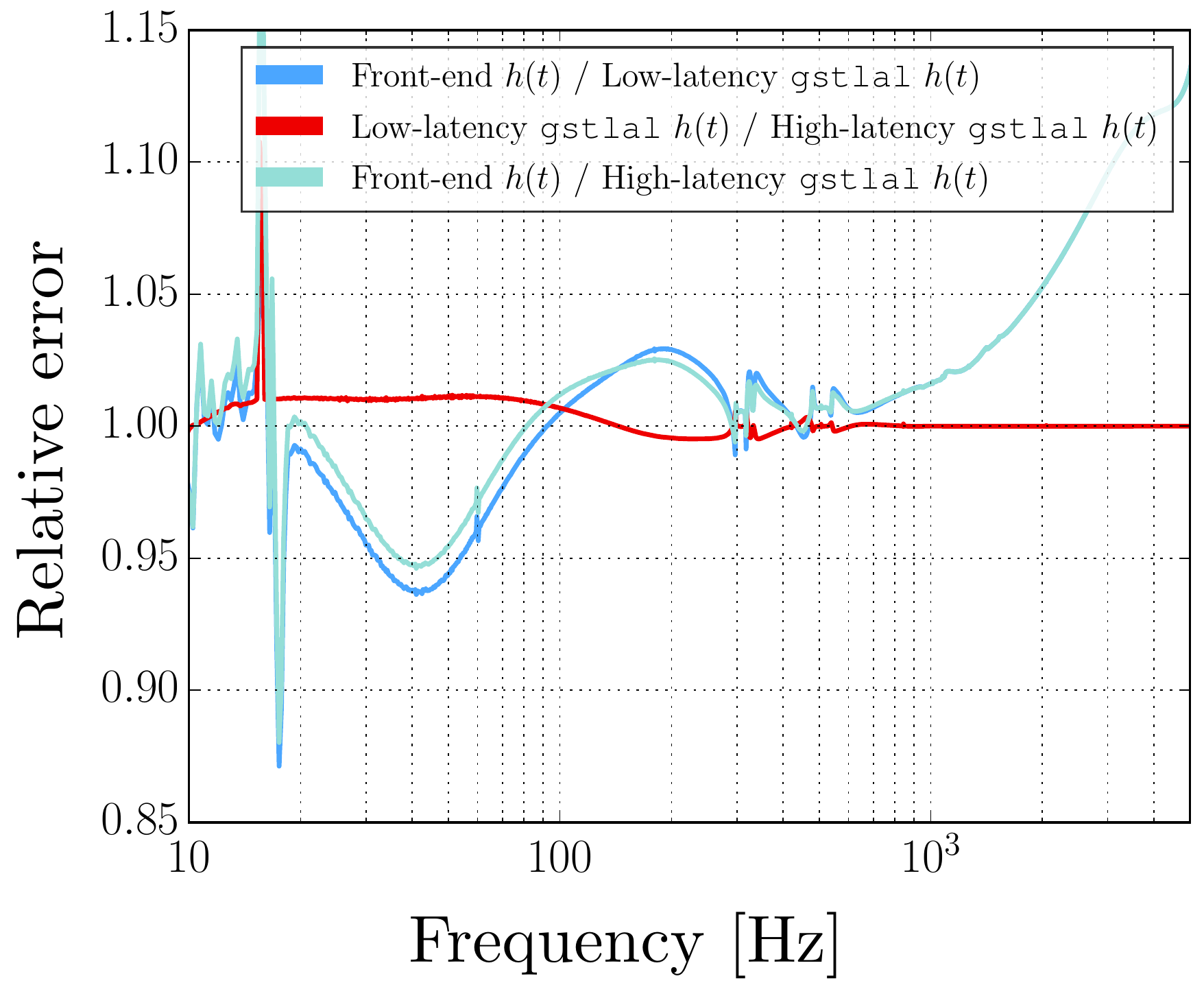}
\end{center}
\caption{A comparison of the output from the front-end calibration pipeline, low-latency \texttt{gstlal} calibration pipeline, and high-latency \texttt{gstlal} calibration pipeline at L1. The plot shows the ratio of the ASDs of the front-end $h(t)$ to the low-latency \texttt{gstlal} $h(t)$ and the ratio of the ASDs of the low-latency \texttt{gstlal} $h(t)$ to the high-latency \texttt{gstlal} $h(t)$.  The calibration undergoes about a 5-10\% improvement between the front-end $h(t)$ and the low-latency \texttt{gstlal} $h(t)$ and an additional few percent improvement between the low-latency \texttt{gstlal} $h(t)$ and the high-latency \texttt{gstlal} $h(t)$.  The biggest cause for the improvement from the front-end $h(t)$ to the low-latency \texttt{gstlal} $h(t)$ is the application of some of the TDCFs in the low-latency \texttt{gstlal} $h(t)$.  The difference between the low- and high- latency \texttt{gstlal} calibration pipeline products is caused primarily by flaws in the front-end pipeline's actuation filters that are uncompensated in the \texttt{gstlal} calibration pipeline.  This is, however, not the case at H1, where the low- and high-latency \texttt{gstlal} calibration pipelines produce nearly identical outputs across the relevant frequency band.}
\label{fig:asdcomparison}
\end{figure*}

\begin{figure*}[t]
\centering
\begin{subfigure}[t]{.49\textwidth}
\centering
\includegraphics[width=\linewidth] {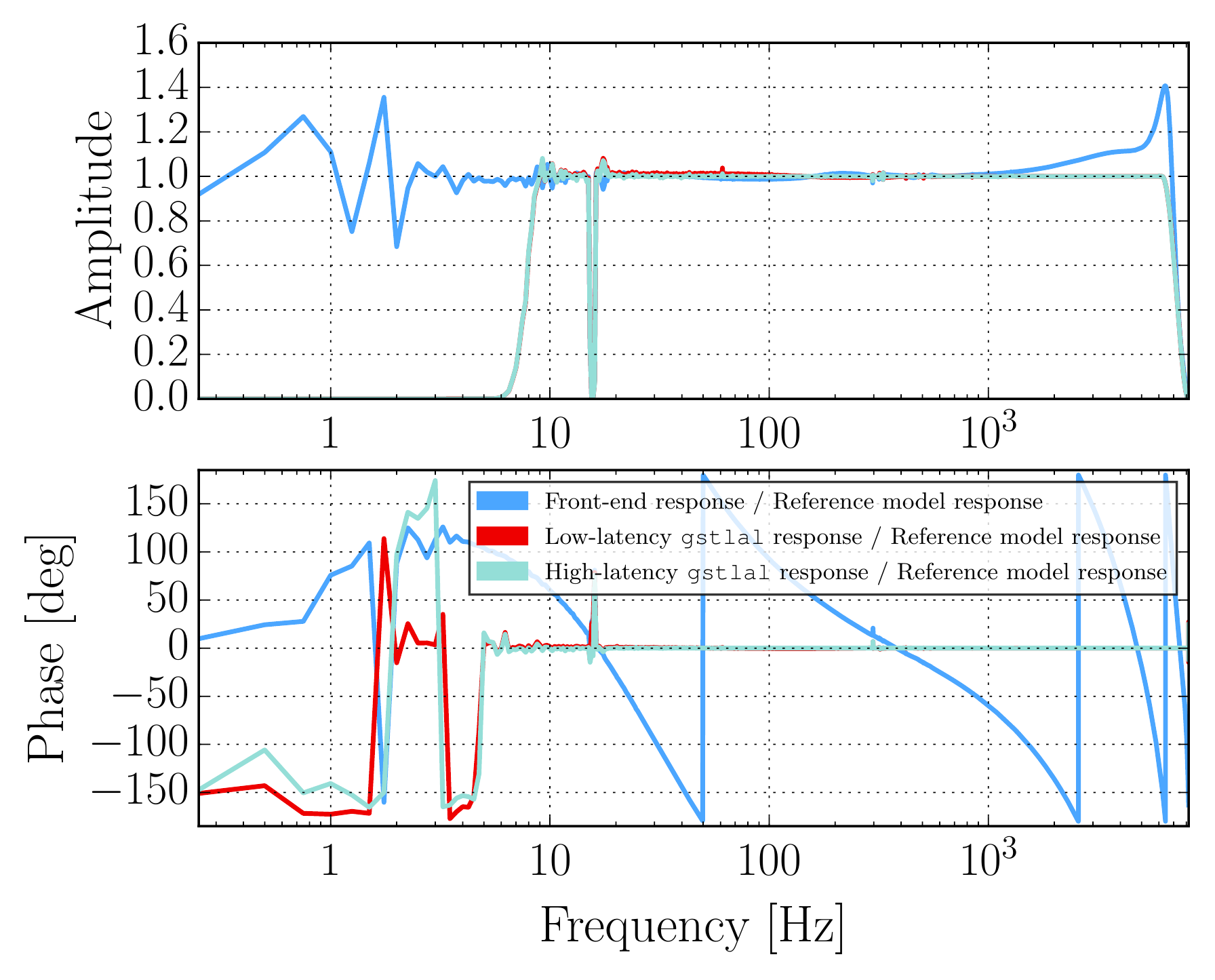}
\end{subfigure}
\begin{subfigure}[t]{.49\textwidth}
\centering
\includegraphics[width=\linewidth] {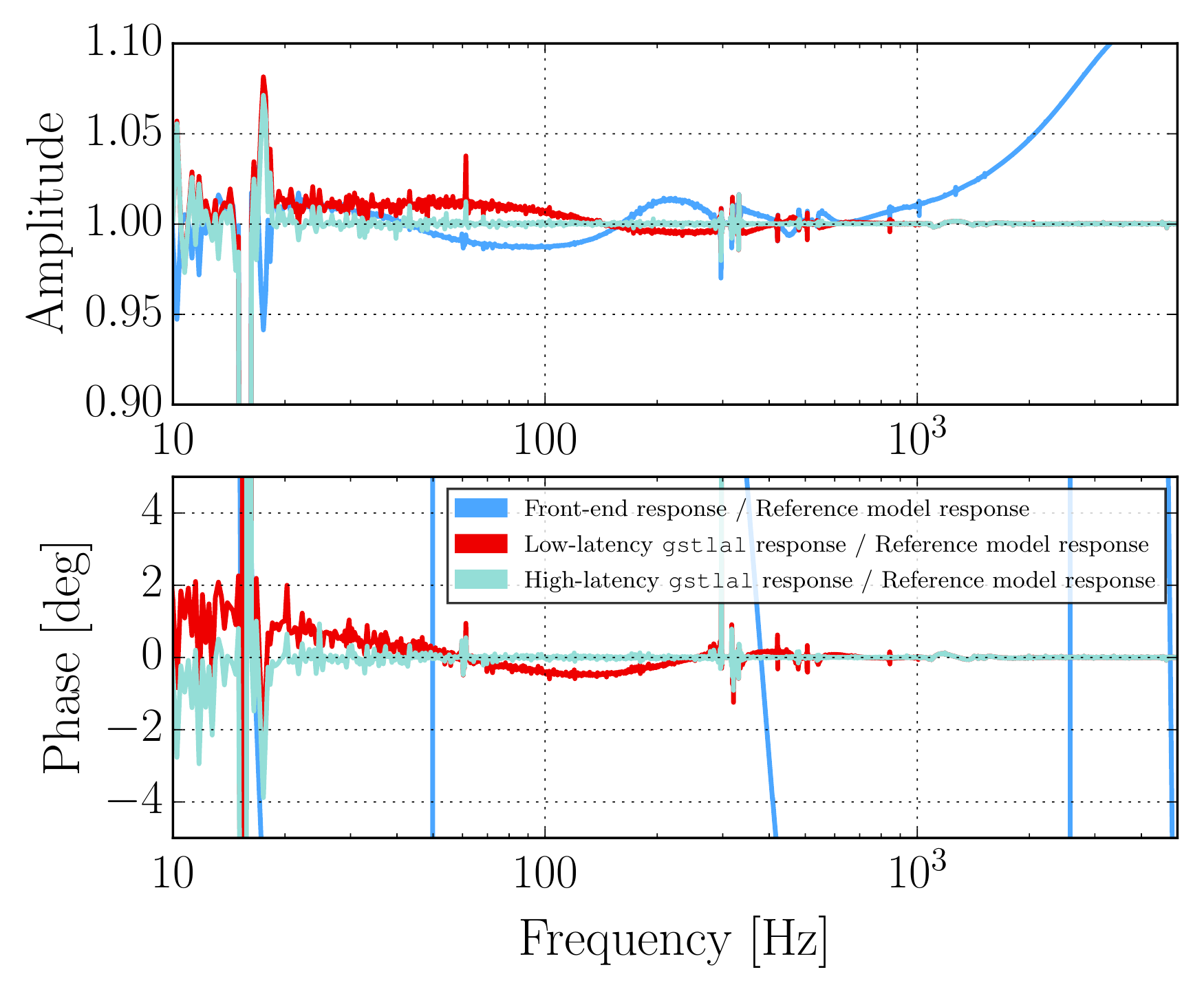}
\end{subfigure}
\caption{A comparison of the response function derived from the output of the front-end calibration pipeline, low-latency \texttt{gstlal} calibration pipeline, and high-latency \texttt{gstlal} calibration pipeline at L1 to the reference model response function.  The plot on the left shows he comparison across all frequencies in order to display the effects of the low- and/or high-pass filters used by each pipeline.  The plot on the right is zoomed in to only display the frequency range [10 Hz, 5 kHz].  This comparison captures the differences in the derived FIR and IIR filter representations of the calibration actuation and sensing functions.  The effects of TDCFs is not included in this comparison.}
\label{fig:responsecomparison}
\end{figure*}

Rapid analysis of data is essential in the era of gravitational-wave astronomy \cite{Singer:2014qca} and a key ingredient to this is providing high-precision calibration as quickly as is possible to the analysis pipelines.  There are three different methods in place for producing calibrated $h(t)$ data during aLIGO's first and second observing runs, each with varying degrees of precision and latency.  The method for producing the lowest latency calibrated data involves using the front-end computer system, which is the system that controls the instrument feedback loops.  However, we do not yet have the software developed to provide a calibrated data product from the front-end that also meets the calibration accuracy requirements for gravitational-wave data analysis activities.  Therefore, a more accurate low-latency calibration is produced using a combination of calibration software written for the front-end computers, known as the front-end calibration pipeline, and a \texttt{GStreamer}-based \cite{gstreamer} pipeline, known as the \texttt{gstlal} calibration pipeline, that improves on the accuracy of the front-end calibration at the cost of adding a few seconds of latency.  

Additionally, there is often a need to recalibrate the data at a later point in time.  The most common reasons for recalibrating the data are dropouts of data somewhere in the low-latency system and improved models for the calibration developed over time.  The high-latency, offline calibration is therefore the most accurate calibrated strain data produced for Advanced LIGO.  The same \texttt{gstlal} calibration pipeline that is used in the low-latency procedure is used for the offline calibration, except it is run in a different mode with a few varying steps and different data and filter inputs.  

Table~\ref{table:pipelines} summarizes the differences between each calibration pipeline.  Fig.~\ref{fig:asdcomparison} shows a comparison of the amplitude spectral density of each calibration pipeline and Fig.~\ref{fig:responsecomparison} shows a comparison of the response function as derived from each calibration pipeline to the reference model response function.  From these figures, it is clear that the improvement from the front-end calibration pipeline to the low-latency \texttt{gstlal} calibration pipeline is significant, while the improvement from the low-latency \texttt{gstlal} calibration pipeline to the high-latency \texttt{gstlal} calibration pipeline is more minor.  While Fig.~\ref{fig:responsecomparison} does illustrate how the response function as derived from each calibration pipeline compares to the reference model response function, this figure does not portray the systematic and statistical uncertainty inherent to the reference model.  For a detailed discussion of the full calibration uncertainty, see \cite{Craig}.

{\it Notation:} Throughout the following sections, quantities computed by the front-end calibration pipeline will be denoted with a superscript 0, quantities computed by the low-latency \texttt{gstlal} calibration pipeline will be denoted with a superscript 1, and quantities computed by the high-latency \texttt{gstlal} calibration pipeline will be denoted with a superscript 2.  This notation also distinguishes the filters used in each pipeline.  For example, $C^0$ is the sensing function IIR filter used in the front-end calibration pipeline and $C^2$ is the sensing function FIR filter used in the high-latency \texttt{gstlal} calibration pipeline.

\subsection{The Front-End Calibration Pipeline } \label{ssec:CALCS}

The front-end calibration pipeline has the advantage of being directly hooked into all of the other front-end computer systems, thereby allowing seamless access to all of the appropriate instrument models and parameters.  This enables the calibration model to remain up-to-date and in-sync with the instrument.  However, the front-end calibration infrastructure is limited by two features common to real-time computer systems.  Originally only delays that were an integer number of digital sampling time intervals, where the digital sampling time interval for a 16384 Hz channel is 61 $\mu$s, could be easily implemented using the front-end system.  This hurdle has been overcome and will be discussed more below.  Additionally, super-Nyquist poles cannot be easily modeled in the front-end calibration, which leads to a systematic error at high frequencies.  The super-Nyquist poles are due to real features of the analog electronics and cannote be modeled in the digital front-end system because the poles are above the Nyquist frequency of the front-end system.


\subsubsection{IIR filters}

The front-end calibration pipeline uses infinite impulse response (IIR) filtering techniques.  Each component of the static reference models for the inverse sensing function and the actuation function are modeled using zero-pole-gain or second-order section IIR filters.  Since it is hard to fit the transfer function measurements with just zero-pole-gain or second-order-section models, these filters introduce additional systematic errors in the calibrated data.

The IIR filters used for the actuation function are exact copies of the digital IIR filters used in the actuation path for interferometer controls whenever possible.  The full suspension model is simplified, reducing the $Q$ factor of some resonances while removing other high-frequency resonances, such as violin modes, before converting from a continuous representation to a discrete, IIR filter representation.  The inverse sensing function is modeled using a single zero, a DC gain, and a high-frequency roll-off.  In order to convert from the continuous representation of the actuation and inverse sensing functions to the discrete representation, we use the built-in \texttt{MATLAB} functions \texttt{c2d} with the bilinear (Tustin) method, which does the continuous to discrete conversion \cite{c2d}, and \texttt{minreal}, which removes very close zero/pole pairs \cite{minreal}.  High-frequency effects in the actuation and inverse sensing paths are not replicated in the front-end calibration pipeline.  

\subsubsection{Front-end calibration pipeline overview}

\begin{figure*}
\begin{center}
\includegraphics[width=0.96\textwidth] {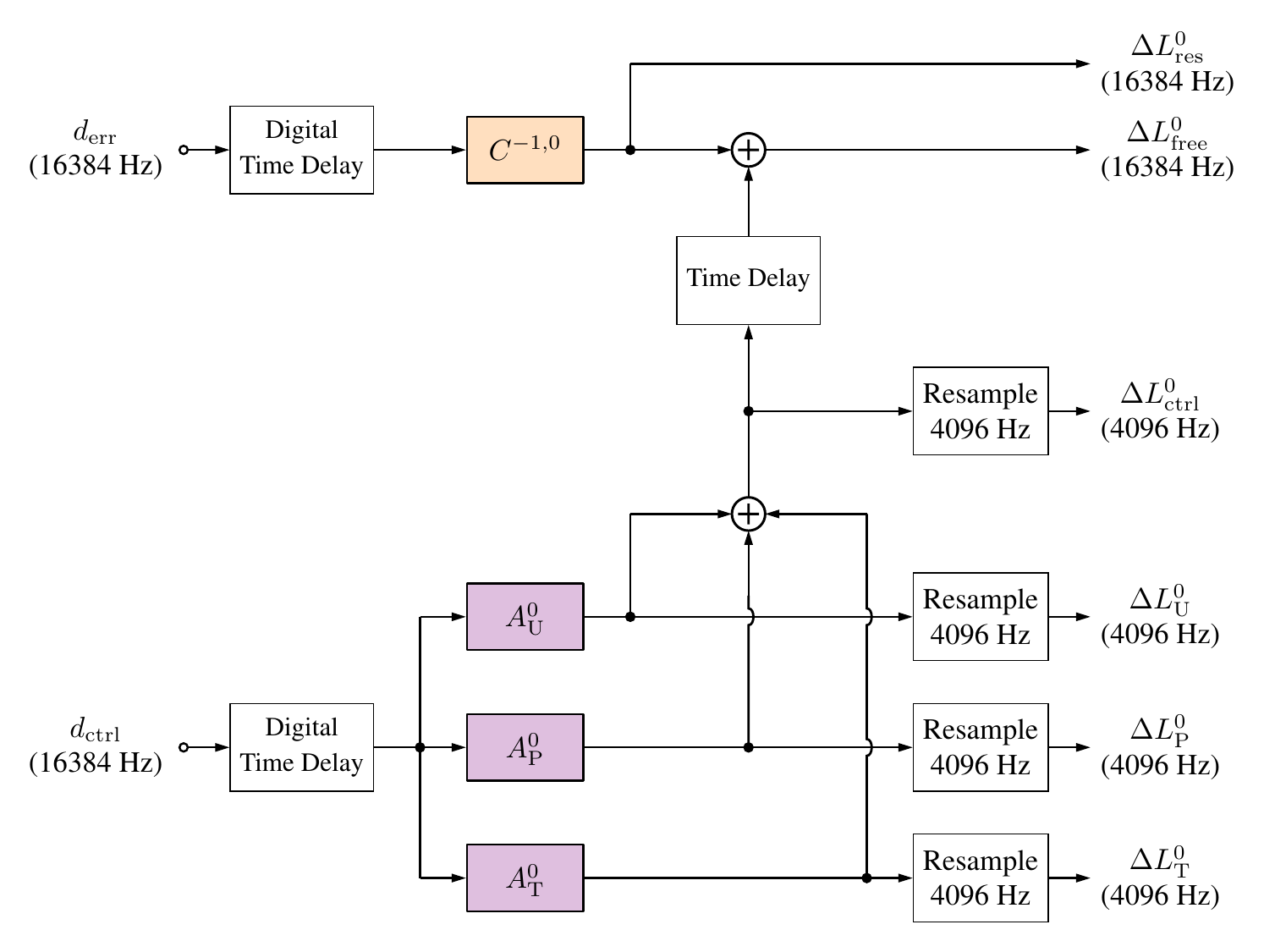}
\end{center}
\caption{Simplified diagram of the front-end calibration pipeline.  The digital error $d_{\rm err}$ and digital control $d_{\rm ctrl}$ signals are picked off from the DARM model in the front-end computers.  This process adds a single digital-sampling-period delay in the $d_{\rm err}$ and $d_{\rm ctrl}$ signals, shown above as the ``Digital Time Delay" box.  The error and control signals are then filtered with the relevant IIR filter representations for the inverse sensing function and the actuation function, respectively.  The control signal path is delayed by an amount representing the net delay between the control and error signal paths before being combined with the error signal to form $\Delta L_{\rm free}^0$.  The filtered control and error signals are also outputted from the pipeline. Details of the calculation of the TDCFs are not shown.}
\label{fig:calcs}
\end{figure*}

The function of the front-end calibration pipeline is to pick-off the signals $d_{\rm err}$ and $d_{\rm ctrl}$ and use Eq.~\eqref{eq:hoft} to calculate the calibrated residual signal $\Delta L_{\rm res}^0$ and the calibrated control signals $\Delta L_{\rm U}^0$, $\Delta L_{\rm P}^0$, and $\Delta L_{\rm T}^0$, where the combined calibrated control signal is $\Delta L_{\rm ctrl}^0 = \Delta L_{\rm U}^0 + \Delta L_{\rm P}^0 + \Delta L_{\rm T}^0$.  The final calibrated output of the front-end calibration is $\Delta L_{\rm free}^0$.  Here we outline the basic steps involved in computing each of these outputs, and these steps are shown pictorially in Fig.~\ref{fig:calcs}.

{\it Pick off error and control signals} Since the error and control digital signals $d_{\rm err}$ and $d_{\rm ctrl}$ are picked off from the DARM front-end model and passed to the front-end calibration model, there is a one digital-sampling-period delay.  

{\it Apply IIR filters} The IIR filters for the inverse sensing function are applied to the $d_{\rm err}$ signal to produce $\Delta L_{\rm res}^0$, which is also saved as an output at a sample rate of 16384 Hz.  The $d_{\rm ctrl}$ signal is split into three equivalent paths.  One path is filtered with the IIR filter models for the upper intermediate actuation function $A_{\rm U}^0$, another one is filtered with the IIR filter models for the penultimate actuation function $A_{\rm P}^0$, and the third one is filtered with the IIR filter models for the test actuation function $A_{\rm T}^0$. 

{\it Resample $\Delta L_{\rm P, U, T}^0$ outputs} After $d_{\rm ctrl}$ is filtered with the three stages of the actuation, the output of each of these filtering processes is resampled from 16384 Hz down to 4096 Hz and saved as the outputs $\Delta L_{\rm P}^0$, $\Delta L_{\rm U}^0$, and $\Delta L_{\rm T}^0$.  This downsampling is done to reduce the amount of storage memory required to save these data.  

{\it Combine and apply delay for $\Delta L_{\rm ctrl}^0$}  The filtered output of the actuation path (before it is resampled down to 4096 Hz) is combined together to form $\Delta L_{\rm ctrl}^0$ at a sample rate of 16384 Hz.  The relative delay between the actuation and sensing paths is then applied to $\Delta L_{\rm ctrl}^0$ to ensure they combine with the correct phase.  This delay requires sub-sample phase accuracy to more accurately represent the difference in the delay along each path and is implemented using a Thiran fractional delay filter \cite{Laakso}.

{\it Compute $\Delta L_{\rm free}^0$} The output $\Delta L_{\rm res}^0$ and $\Delta L_{\rm ctrl}^0$ are added together to form $\Delta L_{\rm free}^0$ at a sample rate of 16384 Hz.  This is the final output of the front-end calibration pipeline.  

\subsubsection{Time-dependent correction factors}

The infrastructure to track the TDCFs is built into the front-end calibration pipeline.  In order to compute the TDCFs, several data streams need to be demodulated at given calibration line frequencies.  The demodulation technique used by the front-end calibration pipeline is to first mix a local oscillator at the given frequency with the incoming signal, then apply a low-pass filter with a corner frequency of 0.1 Hz to the mixed output.  The demodulated lines are used to compute the TDCFs as described in Eqs.~\eqref{eq:kappaT}, \eqref{eq:kappaPU}, \eqref{eq:kappaC}, and \eqref{eq:fcc}. Finally, a 128 second running average is performed using a low-pass filter with a corner frequency of $\approx 0.008$ Hz, similar to the procedure used in the \texttt{gstlal} calibration pipeline, described in Appendix~\ref{appendix:appendixB}.  The second low-pass filter is necessary to avoid sporadic transients that could impact the $h(t)$ data stream if corrections were applied.  The TDCFs are then downsampled to 16 Hz and recorded for comparison with the TDCFs computed in the \texttt{gstlal} calibration pipeline.

The TDCFs computed in the front end are not used to constantly correct the sensing and actuation
functions in any of the calibration pipelines.  This is mainly due to the long response time of the IIR filters used in
the front-end models that suffer from lasting effects of small transients.

\subsubsection{Relative accuracy of the front-end calibration pipeline}

A clear benefit of the front-end calibration process is that calibrated strain data can be produced with sub-second latency.  Since the future of gravitational-wave astronomy relies on very low-latency detection candidate identification, we are working on moving the entire calibration process into the front-end pipeline.  However, work to improve the accuracy of the front-end calibration pipeline to the desired level is still underway.

While the current level of accuracy of $\Delta L_{\rm free}^0$ is acceptable for commissioning purposes, it is relatively poor when compared to the other calibrated data products discussed in the next sections and does not yet meet the standards required for many astrophysical analyses.  Systematic errors are introduced by the IIR filters, and it is challenging to build IIR filters that accurately model super-Nyquist features of the calibration, such as high-frequency poles in the sensing function due to a transimpedance amplifier and analog whitening filters.  We have, however, had good success in modeling these features using FIR filters.  Additionally, we can't apply a phase advance in the real-time code used by the front-end calibration pipeline, and this results in initially recorded data with a fixed and known number of cycles of delay, relative to their actual occurrence in the interferometer.  The sum total of these systematic errors is shown in Fig.~\ref{fig:responsecomparison}, which shows a systematic error as large as 11\% in magnitude at high frequencies and a very large phase error across the frequency band [10 Hz, 5 kHz].

The front-end calibration pipeline currently cannot apply the TDCFs to correct the sensing and actuation functions also due to limitations involved with using IIR filtering.  The systematic error introduced by not applying the TDCFs is evident from Fig.~\ref{fig:asdcomparison} in the frequency range from 20 Hz - 300 Hz as a $\sim$5\% error in magnitude.

\begin{figure*}[t]
	\centering
	\includegraphics[width=0.6\textwidth]{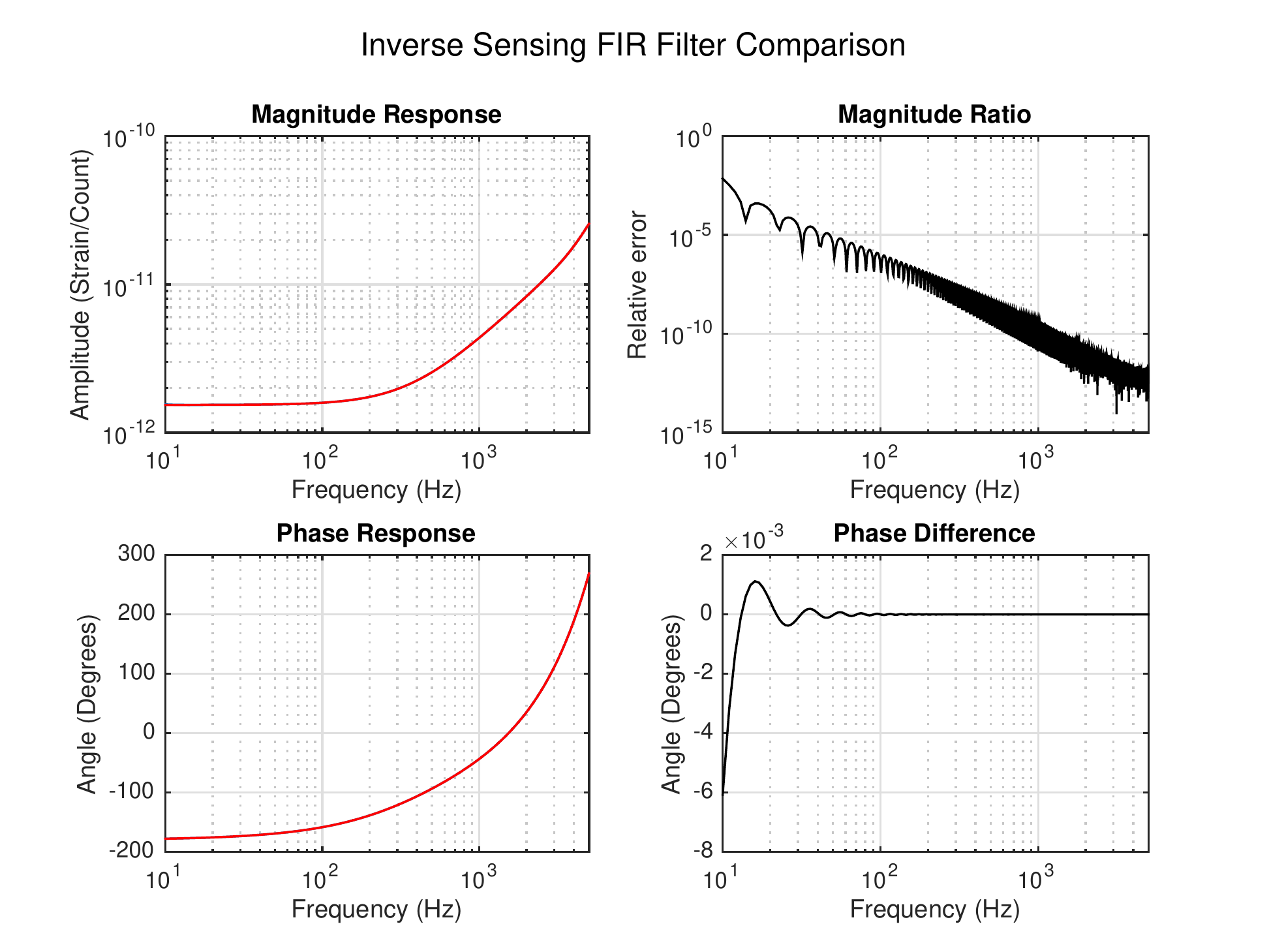}
	\caption{Comparison of the inverse sensing FIR filter to the inverse sensing model.  The left panel shows a bode plot of the frequency response of an example inverse sensing FIR filter for L1 (red) and the original inverse sensing model (blue).  The top right plot shows the relative error in magnitude between the original model and the frequency response of the FIR filter.  The bottom right plot shows the phase difference in degrees between the original model and the frequency response of the FIR filter.  As can be seen in the plots, the FIR filter is a very accurate representation of the derived model for the inverse sensing function above 10 Hz.  (color online) \label{fig:FIRfiltercomparison}}
\end{figure*}

\subsection{The Low-Latency \texttt{gstlal} Calibration Pipeline} \label{ssec:GDS}

Some of the shortfalls of the front-end
calibration procedure that limit the accuracy of the data are compensated for in a calibration
pipeline called the low-latency \texttt{gstlal} calibration pipeline. It takes the output data streams
from the front-end calibration pipeline, and after applying corrections, it produces the calibrated
$h(t)$ time series with latency on the order of several seconds. During O2, the
total latency of the \texttt{gstlal} calibration pipeline was improved from 10-14 seconds
to 5-9 seconds.
The software package used for this second low-latency calibration pipeline is \texttt{gstlal}, a package that wraps LIGO Algorithm Library (\texttt{LAL}) software \cite{lal} with the audio/video streaming software package \texttt{GStreamer} \cite{gstreamer, gstlal}.
The \texttt{gstlal} calibration pipeline applies high-frequency corrections that could not be applied in the front-end calibration, applies the appropriate time delay to each component of the calibration, compensates for the scalar TDCFs and computes a state vector that records the integrity of $h(t)$.  An example release version of the \texttt{gstlal} calibration pipeline can be found in \cite{gstlalcalibration}.

The \texttt{gstlal} calibration pipeline is also used for all high-latency recalibration that  may be required.  Both the low- and high-latency \texttt{gstlal} calibration pipelines use finite impulse response (FIR) filters.  The reasons for using FIR filters instead of IIR filters, which are used by the front-end calibration pipeline, are the following: 1) When recalibrating the data offline, it is useful to recalibrate the data in short, parallelized stretches, which requires FIR filters as opposed to IIR filters;  2) The low-latency calibration system uses both a primary calibration pipeline setup and a redundant pipeline setup to safeguard against online system failures, and in order for the output of the primary and redundant pipelines to be identical, the pipelines cannot depend on their start time or long-term history; 3) All calibration products outside of the front-end need to be reproducible for arbitrarily chosen start times and durations.


\subsubsection{FIR filters} \label{sssec:FIRfilter}

The FIR filters used by the low-latency \texttt{gstlal} calibration pipeline model the super-Nyquist features of the inverse sensing function that are not captured by the front-end calibration pipeline, corrections for systematic errors introduced by the IIR filters used in the front-end calibration pipeline, and accurate time delays in both the sensing and actuation functions.  We do not attempt to correct high-frequency effects in the actuation path, because the actuation function only contributes to the calibration at low frequencies.
The process for creating discrete representations, FIR filters from established continuous representation
models of the sensing and actuation is the following: 

\begin{enumerate}
\item {\it High-pass filter} Seismic noise in the raw data channels is too high at low frequencies to measure any gravitational-wave signals, and the digital system has finite dynamic range. Low frequencies are therefore rolled off by multiplying the frequency components below 9 Hz by half of a Hann window raised to the fourth power.
\item {\it Low-pass filter in the sensing path} Since the inverse sensing function tends toward infinity at high frequencies, a low-pass filter is applied to the inverse sensing function to roll off high frequencies ($f \gtrapprox 6$ kHz) smoothly, in addition to the the high-frequency roll-off in the front end.  This is done by multiplying the high frequency components by half of a Hann window.
\item {\it Artificial delay} An artificial delay is added to the FIR filter that is set to be half of the length of the filter.  This delay is undone within the \texttt{gstlal} calibration pipeline by advancing the filter output by an equivalent number of samples.  The reason for the delay is to center the FIR filter in time, avoiding edge effects while filtering and making the filter non-causal, with output depending on both past and future inputs.
\item {\it Inverse Fourier transform} The Nyquist component is zeroed out and then the inverse Fourier transform is computed to obtain the FIR time-domain filter.
\item {\it Tukey window} A Tukey window function is applied to the resulting time-domain FIR filter to ensure it falls off smoothly at the beginning and end of the time-domain filter response.    All of the above steps are performed using standard \texttt{Matlab} software packages \cite{Matlab}.  
\end{enumerate}

The fidelity of the FIR filters is checked by taking the frequency response of the final FIR filter with the artificial delay removed and comparing the resulting magnitude and phase to the original frequency-domain model.  Example comparison plots are shown in Fig.~\ref{fig:FIRfiltercomparison}.  The differences between the frequency response of the FIR filter and the original frequency domain model are less than \SI{0.1}{\percent} in magnitude and \SI{0.01}{\degree} in phase from 10 Hz to 5 kHz.  Additionally, the fidelity of these FIR filters can be seen by examining Fig.~\ref{fig:responsecomparison}, which shows the combined effect of the FIR filters through the response function.  


\subsubsection{Low-latency \texttt{gstlal} calibration pipeline overview} \label{sssec:GDSpipeline}

\begin{figure*}
\begin{center}
\includegraphics[width=0.96\textwidth] {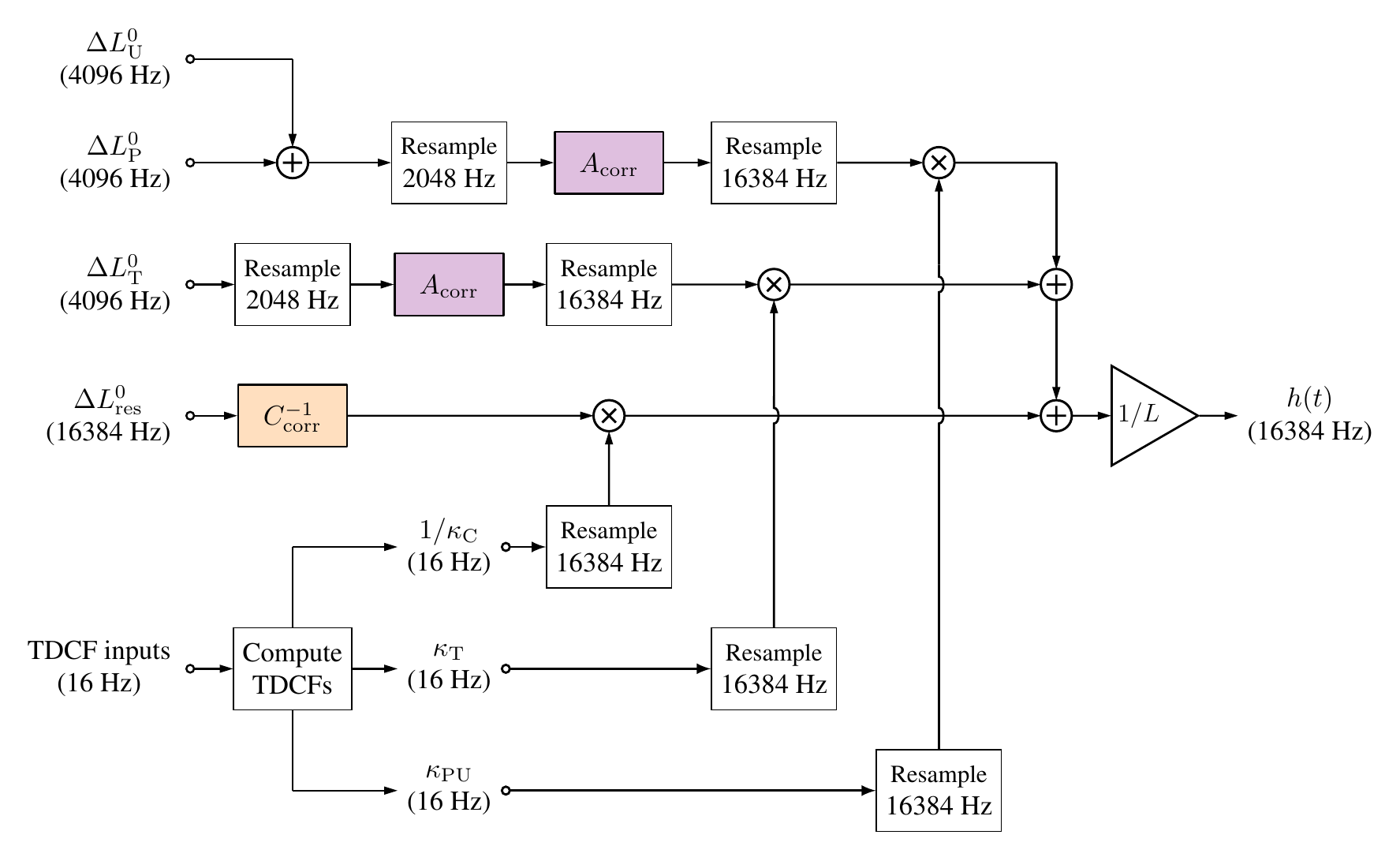}
\end{center}
\caption{Simplified diagram of the low-latency \texttt{gstlal} calibration pipeline. The partially-calibrated outputs of the front-end calibration pipeline are used as inputs and filtered by the appropriate FIR filters for the actuation and inverse sensing paths.  The filtered output is corrected by the TDCFs and added together.  The result is divided by $L$ to give the final results for $h(t)$.  Details of the calculation of the TDCFs and the state vector calculation are not shown.}
\label{fig:GDSpipeline}
\end{figure*}

Rather than reading in the final calibration product of the front-end, $\Delta L_{\rm free}^0$, the \texttt{gstlal} calibration pipeline reads in the components of $\Delta L_{\rm free}^0$ separately. The contribution from the actuation is split into three terms that correspond to displacements form actuators at the three lowest stages of the suspension system: $\Delta L_{\rm T}^0$, $\Delta L_{\rm P}^0$, and $\Delta L_{\rm U}^0$. The contribution from the inverse sensing function is $\Delta L_{\rm res}^0$.  Fig.~\ref{fig:GDSpipeline} shows a simplified diagram of the pipeline.  The following procedure describes the workflow of the low-latency \texttt{gstlal} calibration pipeline chronologically:

{\it Fill in missing data} Sometimes, chunks of data in one or more channels being read in by the \texttt{gstlal} calibration pipeline are missing or corrupted.  These dropouts generally occur with a frequency of a few per week and can range from one second in length to several hours in some cases.  Oftentimes, they occur in all the input channels at once, but this is not always the case. This missing or corrupted data is filled in at the beginning of the pipeline and timestamped appropriately to produce a continuous stream. Most channels are filled in with zeros, except for the coherence uncertainty channels (discussed in Appendix~\ref{appendix:appendixB}), which are filled in with ones. This replacement also occurs when an input sample's magnitude is outside of the expected range of $[10^{-35}, 10^{35}]$, to prevent arithmetic underflows and overflows.

{\it Add $\Delta L_{\rm P}^0$ and $\Delta L_{\rm U}^0$} Since the same FIR filter and time-dependent correction are applied to both, $\Delta L_{\rm P}^0$ and $\Delta L_{\rm U}^0$ can be added before filtering, to save computational cost.  The result is referred to as $\Delta L_{\rm PU}^0$. $\Delta L_{\rm T}^0$ is not combined because it receives a separate time-dependent correction factor $\kappa_T(t)$. 

{\it Downsample the actuation channels} The three actuation channels are read into the \texttt{gstlal} calibration pipeline at a sample rate of 4096 Hz. To keep computational costs manageable, this is downsampled to 2048 Hz before applying the FIR correction filters.  This is necessary for only the actuation channels, due to the longer filter length (discussed more below).  The downsampling is done using the stock \texttt{GStreamer} element \texttt{audioresample} \cite{audioresample}.  A sinc table is used to filter the input before downsampling, in order to minimize aliasing effects.  The sinc table is an acausal filter, centered in time so as to avoid affecting the phase.  It therefore adds a very small latency ($\approx 0.05$ s) to the pipeline.

{\it Apply FIR filters} The FIR correction filters are applied to correct the front-end estimates of the components of DARM:
\begin{subequations}
\begin{align}
    \Delta L_{\rm T}^1(t) &= \left[A_{\rm corr} * \Delta L_{\rm T}^0\right](t), \label{eq:correctionFilterT} \\
    \Delta L_{\rm PU}^1(t) &= \left[A_{\rm corr} * \Delta L_{\rm PU}^0\right](t), \label{eq:correctionFilterPU} \\
    \Delta L_{\rm res}^1(t) &= \left[C_{\rm corr}^{-1} * \Delta L_{\rm res}^0\right](t) \label{eq:correctionFilterRes},
\end{align}
\end{subequations}
where $C_{\rm corr}^{-1}$ is the digital filter applied to correct the front-end estimate of the residual displacement, and $A_{\rm corr}$ corrects the front-end estimate of the controlled length differential. This correction is the same for all three stages of the actuation because the components that cannot be modeled in the front end, such as time delays and anti-imaging filters, are the same for each stage. 

Since the inverse sensing path contains most of the information above \SI{1}{\kilo\hertz} (see Fig.~\ref{fig:spectrumBreakdown}), it is necessary to filter it at the full $h(t)$ sample rate of 16384 Hz.  The inverse sensing correction filter $C_{\rm corr}^{-1}$ is \SI{1}{second} in length and accounts for roughly half of the computational cost of running the pipeline.  The actuation filters are 6 seconds in length, mainly because 
the actuation channels output from the front end have a lot of noise below \SI{10}{\hertz}, which requires a longer filter to attenuate sufficiently.  Due to the length of these filters, it is necessary to filter at \SI{2048}{\hertz} instead of the original actuation sample rate of \SI{4096}{\hertz} provided by the front end.  Since the actuation path contributes a small amount to the total strain above \SI{1}{\kilo\hertz} (see Fig.~\ref{fig:spectrumBreakdown}), this adds a small ($\lessapprox 2$\%) systematic error to $h(t)$ from 1 kHz to 1.5 kHz.


The FIR filtering is performed in the time domain through direct convolution.  The timestamps of the filtered result are then advanced by the appropriate amount to compensate for the artificial delay that was built into the FIR filters. To test timestamp accuracy, the calibration lines injected with the photon calibrator can be recovered from $h(t)$ and compared with the injected sinusoids to ensure the phase is consistent.  This was done with three of the calibration lines at L1, and the measured timestamp differences were negligibly small compared to the sampling periods of the filters. 

The artificial delay built into the FIR filters is half the length of the FIR filter.  Therefore, filtering of the actuation path adds a latency of 3 seconds to the output of the pipeline, due to the 6-second length of the actuation filters. This is a significant fraction of the 5 - 9 s latency of the low-latency \texttt{gstlal} calibration pipeline.  Much of the remaining latency is due to the 4-second length of output $h(t)$ data files, which also accounts for the stated range in latency.


{\it Upsample the actuation channels} 
After filtering at \SI{2048}{\hertz}, $\Delta L_{\rm T}^1(t)$ and $\Delta L_{\rm PU}^1(t)$ are upsampled to the full $h(t)$ sample rate of 16384 Hz.  A sinc table is again used to filter the input, to attenuate frequencies close to the Nyquist frequency.

{\it Apply time-dependent correction factors}
The TDCFs are applied by multiplying the relevant quantities in a sample-by-sample manner. Correcting the filtered output with the TDCFs as a simple multiplication in the time domain is possible because their variation in time is slow compared to the lowest frequencies in the calibrated frequency band.\footnote{Computational cost could be reduced if the TDCFs were applied and the actuation channels summed before filtering, since the control correction filter would only be applied once.  However, due to the excess noise below \SI{10}{\hertz} in the front-end output, multiplying by the TDCFs before high-pass filtering adds significant noise in the relevant frequency band due to noisy fluctuations in the TDCFs.}

{\it Add inverse sensing and actuation paths}
With correction FIR filters and time-dependent corrections applied, the residual displacement is added with the sum of the two controlled displacements to produce the measured free disturbance, $\Delta L_{\rm free}^1$.  This is divided by the average interferometer arm length $L$ to produce the strain $h(t)$.


\subsubsection{Time-dependent correction factors} \label{sssec:kappas}

For the TDCF computation, the \texttt{gstlal} calibration pipeline does not ingest any partially computed output from the front-end calibration pipeline but rather performs the computation from the same raw inputs as used by the front-end calibration pipeline.  Demodulation in the \texttt{gstlal} calibration pipeline is done by multiplying each sample by $e^{-i \omega t}$, where $t$ is the GPS time associated with that sample and $\omega$ is the angular frequency of a calibration line. We then downsample to \SI{16}{\hertz} to keep computational cost manageable and apply a low-pass filter $H$. The result is a complex value representing the amplitude and phase of the calibration line of interest in the signal being measured. For an oscillation with amplitude $a$ and phase angle $\phi$,
\begin{align}
    &H * \left\{e^{-i \omega t} \left[a\,\rm{cos}(\omega \it{t} - \phi) +\it{ n(t)}\right]\right\} = \\
    &H * \left\{\frac{a}{2}\left[e^{-i \phi} + e^{i(-2 \omega t + \phi)}\right] + n(t) e^{-i \omega t} \right\} \approx \frac{a}{2}e^{-i \phi}, \nonumber
\end{align}
where $n(t)$ is noise, that is, anything other than the sinusoidal injection.

Downsampling a time series containing a $\sim$\SI{300}{\hertz} oscillation to \SI{16}{\hertz} would cause aliasing.  Therefore, an anti-aliasing filter is included in the resampling process.\footnote{Due to the need to process complex data, the element used here is different from the one used for resampling the actuation path.}  This effectively forms part of the low-pass filtering process, attenuating the signal to 1\% of its original amplitude at the Nyquist frequency.  The low-pass filter applied after downsampling is a \SI{20} second Hann window applied in the time domain at \SI{16}{\hertz}.  After this, the TDCFs are computed as described in Appendix~\ref{appendix:appendixA}.

Although the calculation of the TDCFs is running constantly, it produces an accurate result only when the interferometer is in a low-noise state, and even then, the resulting time series is quite noisy.  This is largely due to imperfect coherence of the calibration lines in the error signal.  We describe the procedure used to manage this issue in Appendix~\ref{appendix:appendixB}.

$\kappa_{\rm T}$, $\kappa_{\rm PU}$, and $\kappa_{\rm C}$ are used to correct $h(t)$, while the time-dependence of the coupled cavity pole $f_{\rm cc}$ and the optical spring frequency $f_{\rm s}$ and quality $Q$ of the SRC are not corrected for, due to the difficulty of applying frequency-dependent corrections in realtime.  Computed values of $f_{\rm cc}$ can stray as far as 20 Hz from the reference model values and $f_{\rm s}$ varies from about 6 Hz to 9 Hz at H1.

\subsubsection{State vector}

In addition to computing $h(t)$, the \texttt{gstlal} calibration pipeline also computes a bitwise state vector that denotes the integrity of $h(t)$ at a sample rate of 16 Hz.  The top level bits of the state vector are the summary bits used to determine whether or not $h(t)$ should be used for astrophysical analysis.  For more details about the state vector definition during O2, see Appendix~\ref{appendix:appendixC}.

\subsubsection{Relative accuracy of the low-latency \texttt{gstlal} calibration pipeline}

The low-latency \texttt{gstlal} calibration pipeline delivers a relatively accurate $h(t)$ data product with a latency of 5-9 seconds.  As we dive deeper into the era of gravitational-wave astronomy, work continues to improve both the accuracy and the latency of this pipeline.

The static systematic errors currently present in the low-latency \texttt{gstlal} calibration pipeline include errors introduced by downsampling the actuation path and any systematic bias from the FIR filters. Fig.~\ref{fig:responsecomparison} shows that the sum total of these systematic errors are typically less than 5\% in magnitude and a few degrees in phase across the relevant frequency band [20 Hz, 5 kHz], with narrow frequency ranges containing increased systematic error.  Investigations are still underway to understand and improve the systematic error in these narrow frequency regions.  At times, the systematic errors could be much higher than discussed here due to, for example, human errors in the low-latency system.

In addition to the above static systematic errors, the low-latency \texttt{gstlal} calibration pipeline does not correct for the time-changing $f_{\rm cc}$, $f_{\rm s}$, and $Q$ factors.  This systematic error is not assessed in this paper and will be addressed by a future publication.


\subsection{High-latency \texttt{gstlal} Calibration Pipeline} \label{ssec:DCS}

\begin{figure*}
\begin{center}
\includegraphics[width=0.96\textwidth] {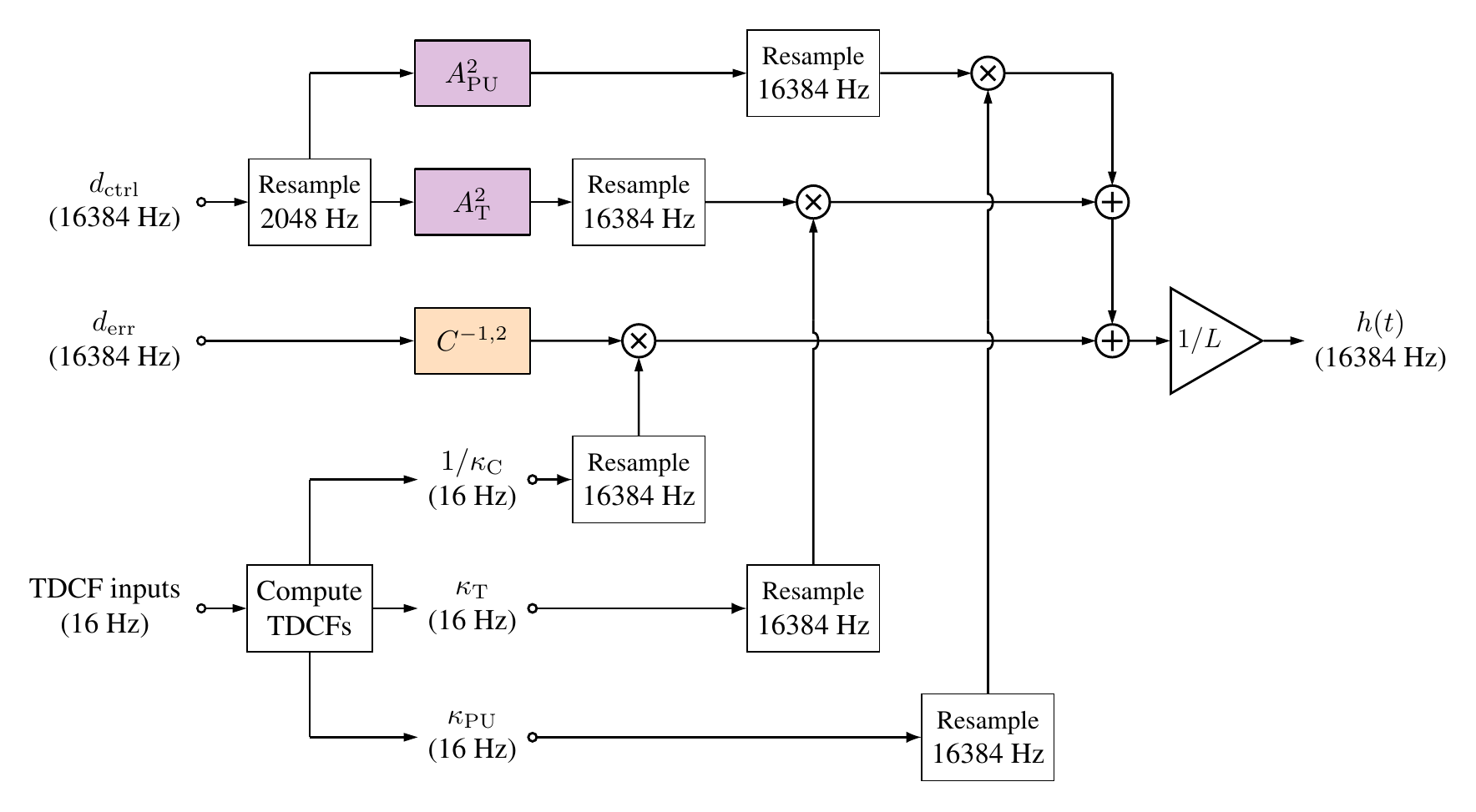} 
\end{center}
\caption{Diagram of the high-latency \texttt{gstlal} calibration pipeline.  The digital error $d_{\rm err}$ and digital control $d_{\rm ctrl}$ signals are calibrated directly using the full inverse sensing and actuation models.  Just as with the low-latency \texttt{gstlal} calibration pipeline, the error and control paths are corrected with the appropriate TDCFs.  A superscript 2 indicates association with the high-latency \texttt{gstlal} calibration pipeline.  Details of the TDCFs calculation and the state vector calculation are not shown.}
\label{fig:DCSpipeline}
\end{figure*}

The \texttt{gstlal} calibration pipeline is also used to produce a second calibration of the archived raw data in high latency.  The high-latency calibration is generally produced weeks to months after acquisition of raw data products.  The \texttt{gstlal} calibration pipeline run in high latency is quite similar to its low-latency counterpart and is depicted in Fig.~\ref{fig:DCSpipeline}.

Instead of reading in the partially-calibrated outputs of the front-end, we calibrate in high latency using $d_{\rm err}$ and $d_{\rm ctrl}$ directly.\footnote{While it would certainly be possible to approach the high-latency calibration with the same methodology as the low-latency calibration, which is to correct the output of the front-end calibration pipeline, it is often impractical to develop the appropriate correction filters that model all the changes occurring in the front-end calibration pipeline during an observing run due to changes caused by human error that inevitably occur in realtime systems.}  Therefore, the FIR filters applied by the high-latency pipeline contain the full static reference models for the inverse sensing and the actuation.  Otherwise, the production of the filters is equivalent to the the description given in Section~\ref{sssec:FIRfilter}.  

In the high-latency calibration, the model-based constants used in computing the TDCFs are recorded in a file which is ingested into the \texttt{gstlal} calibration pipeline.  This is in contrast to these constants being read in directly from the front-end system.  This allows us to correct large systematic errors in the TDCFs caused by mistakes made in low-latency in recording these factors in the front-end system.

The production of high-latency calibration is accomplished by running multiple jobs in parallel on the LIGO Data Grid computing clusters \cite{LDG}.  Each job produces \SI{4096} seconds of calibrated data, using the corresponding stretch of $d_{\rm err}$ and $d_{\rm ctrl}$ with the addition of several minutes of input data at the beginning and end, to allow all filtering processes to settle.

In general, differences between the low-latency and high-latency versions of calibration can vary widely, depending on whether significant systematic errors exist in low-latency that can be corrected in high-latency.  For the majority of O2, however, differences are under \SI{2}{\percent} in magnitude from 10 Hz to 5 kHz, as is seen in Fig.~\ref{fig:asdcomparison}.  The systematic errors that exist in the high-latency \texttt{gstlal} calibration are the same as those in the low-latency \texttt{gstlal} calibration pipeline, with the exception of any human errors that were present in the low-latency process.  Some model improvements are also made between the low- and high-latency calibration iterations, which is the cause of the rough 2\% and 1-3 degree improvement seen in Fig.~\ref{fig:responsecomparison}.

\section{Current and Future Development} \label{sec:futureDevelopment}

During Advanced LIGO's first two observing runs, the low-latency strain $h(t)$ was produced in the time domain using a combination of IIR filtering processes in the front-end computers and FIR filtering processes in the downstream \texttt{gstlal} calibration pipeline.  The \texttt{gstlal} calibration pipeline also applies time-dependent correction factors derived from stable, calibrated reference displacements induced by photon calibrators to the actuation and sensing components of $h(t)$ to improve calibration accuracy in the relevant frequency band.  The \texttt{gstlal} calibration pipeline is used in high latency to produce an additional, more accurate calibration that corrects for missing data and some known systematic errors in the low-latency $h(t)$ data.  During O2, the systematic error and uncertainty achieved by the high-latency calibration (as well as much of the low-latency \texttt{gstlal} calibration) is less than \SI{5}{\percent} in magnitude and \SI{3}{\degree} in phase in the frequency band of \SI{20}{\hertz} to \SI{5}{\kilo \hertz}.  For details of the frequency-dependent uncertainty budget for O2, see \cite{Craig}.

In the future, we hope to improve the latency and accuracy of the low-latency calibration data products.  Work is in progress to implement a method for applying time- and frequency-dependent corrections, i.e., for the time-dependence of the cavity pole $f_{\rm cc}$ and the optical spring frequency $f_{\rm s}$ and quality factor $Q$ of the SRC.  As described in Sec.~\ref{sec:introkappas} and Appendix~\ref{appendix:appendixA}, these are already computed in both the front-end and the \texttt{gstlal} calibration pipeline.  In order to compensate for their time-dependence and improve $h(t)$ accuracy, an algorithm to compute a new inverse sensing filter and smoothly replace the previous filter is being developed.  The time-dependence of the cavity pole is known to be a significant contributor to systematic error in $h(t)$ across the relevant frequency band (see, e.g., Fig.~\ref{fig:pcal_lines}).  The time-dependence of $f_{\rm s}$ and $Q$ is significant at H1 in the lowest frequency band of concern, mainly below about \SI{20}{\hertz}.  An additional change that should lead to a small improvement in the calibration is to compute $\kappa_{\rm P}$ and $\kappa_{\rm U}$ individually instead of tracking and applying their approximate combined effect in the \texttt{gstlal} calibration pipeline as $\kappa_{\rm PU}$.  This would better represent the true model of the actuation as expressed in Eq.~\eqref{eq:A}, and it could have an impact at low frequencies as $\Delta L_{\rm P}$ and $\Delta L_{\rm U}$ differ in their frequency dependence (see Fig.~\ref{fig:spectrumBreakdown}).

The online calibration latency has improved during O2 by about 5 seconds.  This was mainly due to the removal of the latency associated with applying an anti-aliasing filter during the resampling required for the calculation and application of the TDCFs.  Further improvements can be made, primarily in two ways: 1) The length of the gravitational-wave frame files could be reduced from four seconds, which is the current standard, to one second; 2) The length of the actuation filter in the \texttt{gstlal} calibration pipeline can be reduced.  Shortening the actuation filters is straightforward, but if not done with care, spectral leakage will corrupt the $h(t)$ spectrum.  For this reason, we plan to incorporate more effective high-pass filtering into the FIR filters, as the current methodology involves simply rolling off low frequencies with half of a Hann window raised to the fourth power (first step in Section~\ref{sssec:FIRfilter}).

Work is also underway to develop a complete front-end calibration model, as the eventual goal of the LIGO calibration team is to perform the low-latency calibration entirely in the front-end computers.  There are several advantages to having the low-latency calibration entirely in the front end.  The front end's direct access to interferometer models and parameters may reduce the occurrence of human errors in calibration.  Calibrating in the front end also removes data transfer steps, reducing the likelihood of data dropouts.

The primary challenges that must be overcome to realize a complete front-end calibration are the ability to implement FIR filtering and to assign timestamps to data that are not real time.  Due to the length of the FIR filters used in calibration as well as the overhead associated with each 16 kHz front-end computational cycle, the front-end code needs to be modified to distribute the calculations for a single FIR filter over multiple computer cores and spread a single lower rate FIR calculation over multiple 16 kHz cycles.  A fundamental redesign of the interferometer's data acquisition system is needed to be able to advance the calibrated data's timestamp in real-time, as the system currently writes a single 64-second-long frame file at a time with all the data contained having the same timestamp, as well as broadcasting data as it comes in to low-latency pipelines in 1-second-long frame files.

\section{Conclusion} \label{sec:conclusion}

Overall, the calibration process for Advanced LIGO has been expanded to include a very low-latency calibrated strain data product and an unprecedented level of accuracy in the final, high-latency strain data when compared with Initial LIGO.  The latency achieved during O2 run was 5-9 seconds, representing a 5-second improvement made since the beginning of the run.  Typical calibration uncertainty levels in Initial LIGO ranged from \SI{10}{\percent} in magnitude and \SI{10}{\degree} in phase to \SI{20}{\percent} in magnitude and \SI{20}{\degree} in phase \cite{LIGO:2012aa}, while results for the calibration uncertainty around the time of GW170104 \cite{GW170104} are less than \SI{5}{\percent} in magnitude and \SI{3}{\degree} in phase \cite{Craig}.  Future Advanced LIGO observing runs promise further improvement to both the latency and accuracy of the calibration process with an end-goal of having sub-second-latency calibrated strain data with percent-level accuracy available for analysis.  Such an achievement will help push the envelope forward for the new era of gravitational-wave astronomy and multi-messenger astronomy.

\begin{acknowledgments}
The authors would like to thank Chris Wipf, Evan Hall, Jolien Creighton, Patrick Brockill, John Zweizig, Kipp Cannon, Bruce Allen, Chris Pankow, Chad Hanna, and Les Wade for helpful discussions and feedback while developing the methods discussed here.  The authors were supported by National Science Foundation grants PHY-1607178, PHY-1607585, and PHY-1506360. LIGO was constructed by the California Institute of Technology and Massachusetts Institute of Technology with funding from the National Science Foundation and operates under cooperative agreement PHY-0757058 . This paper carries LIGO Document Number LIGO-P1700236.

\end{acknowledgments}

\vfill\null
\appendix
\section{\\Calculation of Time-dependent Correction Factors} \label{appendix:appendixA}

The method used to measure and compensate for temporal variations in the interferometer response (see Sec~\ref{sec:introkappas}) was described in \cite{CALTimeDependence}.  Details of the method are reproduced here for completeness and to show how the additional factors for the detuning of the SRC are included.  

Measurement of the factor $\kappa_{\rm T}$ is done through comparison of a photon calibrator line at frequency $f^{\rm pc}_1$
to another calibration line at frequency $f_{\rm T}$, injected through $x_{\rm T}$ into only the test mass suspension stage.
$\kappa_{\rm PU}$ is measured by comparison of the line at $f^{\rm pc}_1$ to another line at $f_{\rm ctrl}$, which is 
added to the control signal and distributed to each of the lowest three stages in the quadruple pendulum (see Fig.~\ref{fig:DARM_loop}).
All three of these lines are placed between 10 and 40 Hz, where the stages of the actuation
function are comparable in magnitude, to reduce systematic errors.
The quantities $\kappa_{\rm C}$ and $f_{\rm cc}$ are measured using the $f^{\rm pc}_2$ line near 330 Hz,
in the middle of Advanced LIGO's most sensitive frequency band.  The quantities $f_{\rm s}$ and $Q$ are measured by the lowest photon calibrator line $f^{\rm pc}_4$,
placed near \SI{8}{\hertz}, where their effect is significant. The remaining line at $f^{\rm pc}_3$ is used as a check
on calibration uncertainty above 1 kHz.
Calibration lines injected using the electrostatic drive or the photon calibrator can be seen in the $h(t)$ spectrum
(see Fig.~\ref{fig:spectrumBreakdown}).  The DARM line, $x_{\rm ctrl}$, does not appear in $h(t)$ due to the location
where it is injected, before $d_{\rm ctrl}$ is read out.  Table~\ref{tab:callines} summarizes the purpose of each calibration line.

\begin{table}[h!]
\centering
\caption{\label{tab:callines} Summary of the purpose of each calibration line.}
\bigskip
\begin{tabular}{|| c | l | c ||}
\hline 
\textbf{Line}  & \textbf{Purpose} & \textbf{Frequency} \\
\hline
\hline
$f_{\rm ctrl} \rule{0pt}{2.3ex} $ & Computation of $\kappa_{\rm PU}$ & 10 - 40 Hz \\

$f_{\rm T}$ \rule{0pt}{2.3ex} & Computation of $\kappa_{\rm T}$ & 10 - 40 Hz \\

$f^{\rm pc}_1$ \rule{0pt}{2.3ex} & Computation of $\kappa_{\rm T}$ and $\kappa_{\rm PU}$ & 10 - 40 Hz \\

$f^{\rm pc}_2$ \rule{0pt}{2.3ex} & Computation of $\kappa_{\rm C}$  and $f_{\rm cc}$ & $\sim$ 330 Hz \\

$f^{\rm pc}_3$ \rule{0pt}{2.3ex} & Check on high-frequency calibration & $\sim$ 1 kHz \\

$f^{\rm pc}_4$ \rule{0pt}{2.3ex} & Computation of $f_{\rm s}$ and $Q$ & $\sim$ 8 Hz \\
\hline
\end{tabular}
\end{table}

To solve for the TDCFs, we need to know their effect on the amplitude and phase of the calibration lines in $d_{\rm err}$.  From Fig.~\ref{fig:DARM_loop},
\begin{equation}
    \tilde{d}_{\rm err} = \kappa_{\rm C} \frac{\tilde{C}_{\rm res}}{1 + i \, f/f_{\rm cc}} \left(\widetilde{\Delta L}_{\rm free} + \tilde{x}_{\rm pc} - \widetilde{\Delta L}_{\rm ctrl}\right),
\end{equation}
where $\tilde{C}_{\rm res}$ is the static reference model sensing function with dependence on the cavity pole removed.  Note that we have neglected the detuning of the SRC, as this has little impact at the calibration line frequencies used to compute $\kappa_{\rm T}$, $\kappa_{\rm PU}$, $\kappa_{\rm C}$, and  $f_{\rm cc}$.  Referring to Fig.~\ref{fig:DARM_loop}, we see that
\begin{equation} \label{eq:DeltaLCtrlSimple}
    \widetilde{\Delta L}_{\rm ctrl} = \kappa_{\rm T} \tilde{A}_{\rm T} \left(\tilde{D} \, \tilde{d}_{\rm err} + \tilde{x}_{\rm ctrl} - \tilde{x}_{\rm T}\right) + \kappa_{\rm PU} \tilde{A}_{\rm PU} \left(\tilde{D} \, \tilde{d}_{\rm err} + \tilde{x}_{\rm ctrl}\right).
\end{equation}
Solving for $\tilde{d}_{\rm err}$ in terms of the injected excitations, we obtain the expression
\begin{equation}\label{eq:DARMErrWithLines}
    \tilde{d}_{\rm err} = \frac{\widetilde{\Delta L}_{\rm free} + \tilde{x}_{\rm pc} - \kappa_{\rm T} \tilde{A}_{\rm T}\left(\tilde{x}_{\rm ctrl} - \tilde{x}_{\rm T}\right) - \kappa_{\rm PU} \tilde{A}_{\rm PU} \tilde{x}_{\rm ctrl}}{\left(S \, \tilde{C}_{\rm res}\right)^{-1} + \left(\kappa_{\rm T} \tilde{A}_{\rm T} + \kappa_{\rm PU} \tilde{A}_{\rm PU}\right) \tilde{D}},
\end{equation}
where we have defined
\begin{equation}\label{eq:S}
    S \equiv \frac{\kappa_{\rm C}}{1 + i f/f_{\rm cc}}.
\end{equation}
To solve for $\kappa_{\rm T}$, we demodulate $d_{\rm err}$ at the electrostatic drive calibration line $f_{\rm T}$ and the first photon calibrator line $f^{\rm pc}_1$, which are separated by only $\sim$ \SI{1}{\hertz}.  This yields two simple equations:
\begin{equation}\label{eq:DARMErrESDLine}
    \tilde{d}_{\rm err}(f_{\rm T}) = \Biggl.\frac{\kappa_{\rm T} \tilde{A}_{\rm T} \tilde{x}_{\rm T}}{\left(S \, \tilde{C}_{\rm res}\right)^{-1} + \left(\kappa_{\rm T} \tilde{A}_{\rm T} + \kappa_{\rm PU} \tilde{A}_{\rm PU}\right) \tilde{D}} \Biggr|_{f_{\rm T}}
\end{equation}
and
\begin{equation}\label{eq:DARMErrPcalLine}
    \tilde{d}_{\rm err}(f^{\rm pc}_1) = \Biggl.\frac{\tilde{x}_{\rm pc}}{\left(S \, \tilde{C}_{\rm res}\right)^{-1} + \left(\kappa_{\rm T} \tilde{A}_{\rm T} + \kappa_{\rm PU} \tilde{A}_{\rm PU}\right) \tilde{D}}\Biggr|_{f^{\rm pc}_1}\, .
\end{equation}
Then, we take the following ratios of $\tilde{d}_{\rm err}$, $\tilde{x}_{\rm pc}$, and $\tilde{x}_{\rm T}$ at the two line frequencies to estimate $\kappa_{\rm T}$:
\begin{equation}\label{eq:kappaT}
    \kappa_{\rm T} \approx \frac{1}{\tilde{A}_{\rm T}(f_{\rm T})} \cdot \frac{\tilde{d}_{\rm err}(f_{\rm T})}{\tilde{x}_{\rm T}(f_{\rm T})} \left(\frac{\tilde{d}_{\rm err}(f^{\rm pc}_1)}{\tilde{x}_{\rm pc}(f^{\rm pc}_1)}\right)^{-1} \cdot \frac{\tilde{R}^{\rm static}(f_{\rm T})}{\tilde{R}^{\rm static}(f^{\rm pc}_1)}.
\end{equation}
In Eq.~\eqref{eq:kappaT}, we have treated the ratio of the denominators of Eqs.~\eqref{eq:DARMErrESDLine} and \eqref{eq:DARMErrPcalLine} as being constant in time, an approximation that depends on the line frequencies $f^{\rm pc}_1$ and $f_{\rm T}$ being close together. $\tilde{d}_{\rm err}(f_{\rm T})$, $\tilde{x}_{\rm T}(f_{\rm T})$, $\tilde{d}_{\rm err}(f^{\rm pc}_1)$, and $\tilde{x}_{\rm pc}(f^{\rm pc}_1)$ are constantly measured by the calibration pipelines, while the complex constant $\tilde{R}^{\rm static}(f_{\rm T}) / \left(\tilde{A}_{\rm T}(f_{\rm T}) \tilde{R}^{\rm static}(f^{\rm pc}_1)\right)$ depends only on static functions at the line frequencies obtained from measurements, and is therefore read into the calibration pipelines, along with several other constants used in computing the TDCFs.

A similar method is used to compute $\kappa_{\rm PU}$ with the DARM line $f_{\rm ctrl}$ and the same photon calibrator line $f^{\rm pc}_1$, yielding the result
\begin{align}\label{eq:kappaPU}
    \kappa_{\rm PU} \approx -\frac{1}{\tilde{A}_{\rm PU}(f_{\rm ctrl})} \Biggl[\frac{\tilde{d}_{\rm err}(f_{\rm ctrl})}{\tilde{x}_{\rm ctrl}(f_{\rm ctrl})} \left(\frac{\tilde{d}_{\rm err}(f^{\rm pc}_1)}{\tilde{x}_{\rm pc}(f^{\rm pc}_1)}\right)^{-1} &\cdot \frac{\tilde{R}^{\rm static}(f_{\rm ctrl})}{\tilde{R}^{\rm static}(f^{\rm pc}_1)} \Biggr. \\
                                 + \, \Biggl.\kappa_{\rm T} & \tilde{A}_{\rm T}(f_{\rm ctrl})\Biggr]. \nonumber
\end{align}
As seen in Eq.~\eqref{eq:kappaPU}, $\kappa_{\rm PU}$ depends on $\kappa_{\rm T}$, which is computed first.  The optical gain correction factor $\kappa_{\rm C}$ and the cavity pole $f_{\rm cc}$ depend on both $\kappa_{\rm T}$ and $\kappa_{\rm PU}$ under this formalism.  To measure them, we make use of a higher calibration line injected with the photon calibrator, $f_2^{\rm pc} \sim \SI{330}{\hertz}$, where the first term in the denominator of Equation~\ref{eq:DARMErrWithLines} dominates.  Solving for $S(f^{\rm pc}_2)$, we have
\begin{equation}
    S(f^{\rm pc}_2) = \Biggl.\frac{1}{\tilde{C}_{\rm res}} \Biggl(\frac{\tilde{x}_{\rm pc}}{\tilde{d}_{\rm err}} - \tilde{D}\left[\kappa_{\rm T}\tilde{A}_{\rm T} + \kappa_{\rm PU}\tilde{A}_{\rm PU}\right]\Biggr)^{-1} \Biggr|_{f^{\rm pc}_2}.
\end{equation}
Then the expressions for $\kappa_{\rm C}$ and $f_{\rm cc}$ are simply
\begin{eqnarray}\label{eq:kappaC}
    \kappa_{\rm C} &=& \frac{\left|S(f^{\rm pc}_2)\right|^2}{\Re \left[S(f^{\rm pc}_2)\right]}, \\
    \label{eq:fcc}
        f_{\rm cc} &=& -\frac{\Re \left[S(f^{\rm pc}_2)\right]}{\Im \left[S(f^{\rm pc}_2)\right]} f_2^{\rm pc}.
\end{eqnarray}

In order to compute the optical spring frequency $f_{\rm s}$ and quality factor $Q$ of the SRC, we need to consider the full model of the sensing function described in Eq.~\eqref{eq:C}.  In a similar fashion to the derivation above, we express $d_{\rm err}$ in terms of the injected excitations and the TDCFs, this time including the full sensing function model:
\begin{align}\label{eq:DerrFull}
    \tilde{d}_{\rm err} &= \frac{f^2}{f^2 + f_{\rm s}^2 - i f f_{\rm s} / Q} \frac{\kappa_{\rm C} \tilde{C}_{\rm res}}{1 + i \, f/f_{\rm cc}} \left(\tilde{\Delta L}_{\rm free} + \tilde{x}_{\rm pc} - \tilde{\Delta L}_{\rm ctrl}\right) \nonumber \\[10 pt]
                        &= \frac{\tilde{\Delta L}_{\rm free} + \tilde{x}_{\rm pc} - \kappa_{\rm T} \tilde{A}_{\rm T}\left(\tilde{x}_{\rm ctrl} - \tilde{x}_{\rm T}\right) - \kappa_{\rm PU} \tilde{A}_{\rm PU} \tilde{x}_{\rm ctrl}}{(1 + \xi)(S \, \tilde{C}_{\rm res})^{-1} + \left(\kappa_{\rm T} \tilde{A}_{\rm T} + \kappa_{\rm PU} \tilde{A}_{\rm PU}\right) \tilde{D}},
\end{align}
where we have defined
\begin{equation}
    \xi \equiv \frac{f_{\rm s}^2 - i f f_{\rm s} / Q}{f^2}
\end{equation}
and again used Eq.~\eqref{eq:DeltaLCtrlSimple} to substitute for $\tilde{\Delta L}_{\rm ctrl}$.  Demodulating $d_{\rm err}$ at the lowest photon calibrator line ($f^{\rm pc}_4 = \SI{7.93}{\hertz}$ at LHO) yields the expression
\begin{equation}
    \tilde{d}_{\rm err}(f^{\rm pc}_4) = \Biggl. \frac{\tilde{x}_{\rm pc}}{\left(1 + \xi \right) \left(S \tilde{C}_{\rm res}\right)^{-1} + \left(\kappa_{\rm T} \tilde{A}_{\rm T} + \kappa_{\rm PU} \tilde{A}_{\rm PU}\right) \tilde{D}} \, \Biggr|_{f^{pc}_4}.
\end{equation}
Therefore, $\xi(f^{\rm pc}_4)$ can be computed in the \texttt{gstlal} calibration pipeline using
\begin{equation}
    \xi(f^{\rm pc}_4) = -1 + \Biggl[S \, \tilde{C}_{\rm res} \left(\frac{\tilde{x}_{\rm pc}}{\tilde{d}_{\rm err}} - (\kappa_{\rm T} \tilde{A}_{\rm T} + \kappa_{\rm PU} \tilde{A}_{\rm PU}) \tilde{D} \right)\Biggr]_{f^{\rm pc}_4},
\end{equation}
and $f_{\rm s}$ and $Q$ take the simple form
\begin{subequations}
\begin{align}
    f_{\rm s} = f^{\rm pc}_4 \sqrt{\Re[\xi(f^{\rm pc}_4)]}\, , \\
    Q = - \frac{\sqrt{\Re[\xi(f^{\rm pc}_4)]}}{\Im[\xi(f^{\rm pc}_4)]}\, .
\end{align}
\end{subequations}

Note that the calculation of $f_{\rm s}$ and $Q$ depends on the other TDCFs, which are computed first.  Here, we have assumed that the effect of SRC detuning is very small at the higher frequencies used to compute $\kappa_{\rm T}$, $\kappa_{\rm PU}$, $\kappa_{\rm C}$, and $f_{\rm cc}$.  This is valid as long as $f_{\rm s}$ is small compared to those frequencies.  We have not neglected the effect of the previously computed TDCFs on $f_{\rm s}$ and $Q$, as the optical gain and actuation strength clearly have significant contributions here.

\section{\\Smoothing of Time-dependent Correction Factors} \label{appendix:appendixB}

As noted in Section~\ref{sssec:kappas}, the computed TDCFs cannot be applied to $h(t)$ directly due to both excessive noise and the fact that the measurements of the calibration lines are inaccurate when detector noise increases.  Here, we describe the method used to attenuate noise in the TDCFs and to discriminate between times of acceptable and unacceptable measurements. 

The first step is to accept or reject computed TDCFs based on the coherence between the injection channels $x_i$ and error signal $d_{\rm err}$ at the calibration line frequencies.  To compute the coherence between two signals $x(t)$ and $y(t)$ at a frequency $f$, the signals are demodulated at the chosen frequency using a local oscillator.  Then, the coherence is computed as
\begin{equation}
    \gamma^2_{\rm xy}(f) = \frac{| \langle \tilde{x}^*(f) \tilde{y}(f) \rangle |^2}{\langle |\tilde{x}(f)|^2 \rangle \langle |\tilde{y}(f)|^2 \rangle},
\end{equation}
where the angled brackets denote averages, computed using a low-pass filter, and the superscript asterisk denotes complex conjugation.  Coherences are computed using 10-second chunks of input data, and $n_{\rm d} = 13$ independent, consecutive values are then averaged, so that each averaged coherence is based on 130 seconds of input data.  The average coherence is used to compute the normalized random error in the magnitude of the transfer function $\hat H_{\rm x y}$ at each calibration line frequency:
\begin{equation}
    \epsilon \left[\left|\hat{H}_{\rm xy}\right|\right] \approx \sqrt{\frac{1 - \gamma^2_{\rm xy}}{2 n_{\rm d} \gamma^2_{\rm xy}}}
\end{equation}
as derived in \cite{BendatPiersolCoherenceUncertainty}.  This uncertainty is computed in the front end and read into the \texttt{gstlal} calibration pipeline.  If the estimated uncertainty is below the chosen threshold (currently 0.004 at LHO and 0.02 at LLO), a computed TDCF is accepted and passed downstream; otherwise, it is flagged as a gap.  Each TDCF is accepted or rejected based on the estimated uncertainty at the frequency of every calibration line used in its computation.  Very roughly speaking, the coherence of the calibration lines is measured to be acceptable about half of the time during an observing run, although this varies widely with time.

\begin{table*}
\caption{\label{tab:statevector} A summary of the meaning of each bit in the calibration state vector.}
\centering
\begin{tabular} { || c | c | c ||}
\hline
\textbf{bit} & \textbf{Short descriptor} & \textbf{Long descriptor} \\
\hline
\hline
0 & HOFT\_OK & $h(t)$ was successfully computed (logical AND of bits 2-4, 11, 13, 17, and 25)\\
1 & OBS\_INTENT & interferometer is in ``observation intent" mode \\
2 & OBS\_READY & interferometer is in ``observation ready" mode \\
3 & HOFT\_PROD & $h(t)$ was produced \\
4 & FILTERS\_OK & calibration filters settled in \\
5 & NO\_STOCH\_HW\_INJ & No stochastic hardware injections present \\
6 & NO\_CBC\_HW\_INJ & No compact binary coalescence hardware injections present \\
7 & NO\_BURST\_HW\_INJ & No burst hardware injections present \\
8 & NO\_DETCHAR\_HW\_INJ & No detector characterization hardware injections present \\
9 & NO\_GAP & The input data was present (not a gap) \\
10 & KAPPA\_SMOOTHING\_OK & TDCFs smoothing algorithm is settled in \\
11 & KAPPA\_TST\_SMOOTH\_OK & Smoothed $\kappa_{\rm T}$ output is in expected range \\
12 & KAPPA\_TST\_MEDIAN\_OK & Median array used for $\kappa_{\rm T}$ smoothing not dominated by bad coherence time \\
13 & KAPPA\_PU\_SMOOTH\_OK & Smoothed $\kappa_{\rm PU}$ output is in expected range \\
14 & KAPPA\_PU\_MEDIAN\_OK & Median array used for $\kappa_{\rm PU}$ smoothing not dominated by bad coherence time \\
15 & not in use & not in use \\
16 & not in use & not in use \\
17 & KAPPA\_C\_SMOOTH\_OK & Smoothed $\kappa_{\rm C}$ output is in expected range \\
18 & KAPPA\_C\_MEDIAN\_OK & Median array used for $\kappa_{\rm C}$ smoothing not dominated by bad coherence time \\
19 & F\_CC\_SMOOTH\_OK & Smoothed $f_{\rm cc}$ output is in expected range \\
20 & F\_CC\_MEDIAN\_OK & Median array used for $f_{\rm cc}$ smoothing not dominated by bad coherence time \\
21 & SUS\_COH\_OK & Coherence uncertainty for SUS calibration line is acceptable \\
22 & DARM\_COH\_OK & Coherence for DARM calibration line is acceptable \\
23 & PCALY\_LINE1\_COH\_OK & Coherence for first photon calibrator line is acceptable \\
24 & PCALY\_LINE2\_COH\_OK & Coherence for second photon calibrator line is acceptable \\
25 & NO\_UNDERFLOW\_INPUT & Magnitude of input for all channels is between $1\times 10^{-35}$ and $1\times 10^{35}$ \\
26 & F\_S\_SMOOTH\_OK & Smoothed $f_{\rm s}$ output is in expected range \\
27 & F\_S\_MEDIAN\_OK & Median array used for $f_{\rm s}$ smoothing is not dominated by bad coherence time \\
28 & Q\_SMOOTH\_OK & Smoothed $Q$ of SRC output is in expected range \\
29 & Q\_MEDIAN\_OK & Median array used for $Q$ smoothing is not dominated by bad coherence time \\
\hline
\end{tabular}
\end{table*}

The accepted values are then passed to a \SI{128}{\second} running median, where each new value replaces the oldest value currently in the median array.  If computed values are being rejected due to unacceptable coherence, the algorithm will recognize the gap flag associated with those values and instead replace the oldest value in the array with the previously computed median.  Thus, when detector noise increases, the reported TDCFs stabilize at the most recent median value computed during the low-noise state.  The choice to use a running median instead of, e.g., a running average, is intended to make the result insensitive to occasional outliers that occur in the computed TDCFs, resulting in a smoother time series.  The drawback is that computing a median over \SI{128}{\second} of data can be computationally expensive.  To offset this extra cost, we have taken advantage of the fact that, in a running median, the \makebox[\linewidth][s]{previous median is known.  This can be used to reduce the}
\FloatBarrier
\noindent number of operations from $\mathcal{O}(\ell^2)$ to $\mathcal{O}(\ell)$, where $\ell$ is the length of the median array.

A short, 10 second running average is used after the running median in order to remove ``kinks" left after the running median.  This is computationally cheap and reduces the high-frequency content of the smooth TDCF time series.

All of the gating and smoothing operations are carried out at \SI{16}{\hertz}, before the TDCFs are upsampled to \SI{16384}{\hertz} and used to correct $h(t)$.

\section{\\Calibration State Vector} \label{appendix:appendixC}

The \texttt{gstlal} calibration pipeline computes a bitwise state vector that summarizes the integrity of the calibration at a sample rate of 16 Hz.  The definition of each bit in the calibration state vector during O2 is shown in Table~\ref{tab:statevector}.  A value of zero for a bit means the bit is False and a value of 1 means the bit is True.

The zeroth bit of the state vector indicates that $h(t)$ is considered valid and usable for astrophysical analyses.  Bit 1 in the state vector is determined based on state information provided by the operator and the front-end system.  If bit 1 is set to True, then the operator has determined that no commissioning activities are ongoing and science-quality data should be available.  Similarly, bit 2 is determined based on state information provided by the front-end system and indicates that the interferometer has reached a nominal low-noise configuration.  Bit 3 indicates whether or not strain was computed by the pipeline, but this bit does not necessarily indicate the validity of the strain data at that time.

Bit 4 indicates whether the various FIR filters have settled in since the interferometer reached a nominal low-noise state as determined by bit 2.  This bit will be turned off for $N$ seconds after the observation-ready state is reached and will also be turned off for $N$ seconds before the observation-ready state is lost, where $N$ seconds is the total filter settle time for all FIR filters in the pipeline.  This is due to the non-causal nature of the FIR filters, as discussed in Sec.~\ref{sssec:FIRfilter}.

Bits 5-8 indicate whether any known hardware injections are present in the data.  A hardware injection is achieved by actuating the end test masses of the interferometers to mimic an astrophysical signal.  Hardware injections are performed for both transient and continuous gravitational-wave emission \cite{Biwer:2016oyg}.  Bit 9 indicates whether there are times that no input data was received by the calibration pipeline.  Bits 10-24 and 26-29 are all associated with the TDCFs calculation and its smoothing process (see Appendix~\ref{appendix:appendixA}).  In particular, bits 11, 13, and 17 indicate whether or not the computed TDCFs that are used to correct $h(t)$ ($\kappa_{\rm C}$, $\kappa_{\rm PU}$, and $\kappa_{\rm T}$) are within some expected range of values.  If the TDCFs stray outside of the expected range, this is a red flag that something about the current reference calibration model may now be drastically inaccurate.  Bit 25 indicates whether the input could have caused arithmetic underflows or overflows.

Data analysts are advised to use the logical AND of bits 0 and 1 to determine whether to analyze data for astrophysical signals.  Bit 0 indicates the overall integrity of the $h(t)$ calculation while bit 1 indicates that the operator has determined the interferometer is in trustworthy, science-quality data-taking state.


\bibliographystyle{apsrev4-1}
\bibliography{aligo_hoft_paper}

\end{document}